\documentclass[12pt,a4paper,jhep]{article}
\pdfoutput = 1
 \usepackage{jheppub}
\usepackage{textcomp}
\usepackage{subfloat}
{
\usepackage{graphics}
\usepackage{comment}
\usepackage{graphicx} 
\textwidth 6.3in
\newcommand{\be}{\begin{equation}}
\newcommand{\ee}{\end{equation}} 
\newcommand{\ba}{\begin{array}}
\newcommand{\ea}{\end{array}}
\newcommand{\bea}{\begin{eqnarray}}
\newcommand{\eea}{\end{eqnarray}}

\usepackage[utf8]{inputenc}
\usepackage{amsmath}
\usepackage{amsfonts}
\usepackage{amssymb}
\usepackage{makeidx}
\usepackage{multirow}
\usepackage{gensymb}
\usepackage{verbatim}
\usepackage{hyperref}
\usepackage{lmodern}
\usepackage{slashed}
\usepackage{hhline}
\usepackage{mdwlist}
\usepackage{color}
\usepackage[T1]{fontenc}
\usepackage{cleveref}
\usepackage{multicol}
\usepackage{xcolor}
\usepackage{appendix}
\usepackage{caption}
\usepackage{subcaption}

\title{  Momentum dependent gap   in holographic superconductors revisited}
\author[a]{Debabrata Ghorai,}
\author[b]{Yoon-Seok Choun,}
\author[a]{Sang-Jin Sin,}
\emailAdd{dghorai123@gmail.com} 
\emailAdd{ychoun@gmail.com}
\emailAdd{sangjin.sin@gmail.com}
\affiliation[a]{ Department of Physics, Hanyang University, Seoul 04763, South Korea }
\affiliation[b]{Asia Pacific Center for Theoretical Physics (APCTP), Pohang 790-784, South Korea}

\abstract{ 
We reconsider the angular dependence in gap structure  of   holographic superconductors, which has not been treated carefully so far. 
 For the  vector field model, we  show that the normalizable  ground state is  in the  p-wave state because s-wave state is not normalizable. 
 On the other hand,   in the scalar order model, the ground state is in the $s$-wave.  
 The angle dependent gap function is explicitly constructed  in these models.  We also suggest  the modified ansatz of the vector order   which enables to discuss the order $p_{x}\pm ip_{y}$ gap. 
 We have also analytically investigated  the  critical temperature and the behavior of the gap near there.  
  Interestingly, for the fixed conformal dimension of the Cooper pair operator,  {\it the critical temperature in vector model is higher than that of    the scalar  model.} }	
 
\subheader{\today }

\begin{document}
\maketitle

\section{Introduction}
The gauge/gravity duality \cite{adscft1}-\cite{adscft4} has been applied  to strongly correlated systems  as an efficient  tool to handle the strong coupling.
In the context of this duality,
 a theory of superconductivity was set up  \cite{hs6a} using  the spontaneous symmetry breaking of the $U(1)$ symmetry  of an Abelian Higgs model coupled to $AdS$ the gravity \cite{hs1},\cite{hs2},  after which huge number of  investigations on $s$-wave holographic superconductors \cite{shsc1}-\cite{shsc2} has been reported in past decade. 
Although  the  original  model   \cite{hs6a} allowed to estimate the gap and the critical temperature of conductivity of isotropic system, 
typical high $T_c$ superconductors  
show the momentum dependent  gap  structure $\Delta_{k}$ \cite{ cuexpt1,cuexpt2,cuexpt4} which  has been
considered as one of the most important finger prints  of high $T_c$ superconductors.    To address $p$-wave superconductors, Gubser \cite{gubserp} first introduced the non-Abelian gauge field  for holographic superconductor model, which is followed by  many investigations on $p$-wave \cite{phsc1}-\cite{phsc2} or $d$-wave \cite{dhsc1}-\cite{dhsc2} holographic superconductors using abelian  vector and tensor fields. However, to our surprise, none of  the  investigations addressed the momentum $\vec{k}$ dependence of the  superconducting gap, because  in all the previous  works, 
non-vanishing components of vector $A_{\nu}$  or tensors $B_{\mu\nu}$ were assumed   to be isotropic.     Although the angle dependence of the gap was introduced  in a notably exceptional paper  \cite{dhsc1},   the   angle  dependence in that work was introduced by considering the   `fermion spectrum' explicitly rather than through the gap equation, that is,  the equation of motion  of the  complex scalar  function. For more review, see  \cite{ghsc1}-\cite{ghsc3}.

In this paper, we  study the momentum dependence of the gap  function  in holographic set-up.  We consider angle dependent fields in two holographic models, namely, scalar field model and vector field model. We first consider the vector field model in order to  see whether  the role of  the  spin-$1$ field model is essential. We have explicitly shown that the normalized ground state of the vector model is provided only by the $p$-wave state and the s-wave state is not normalizable.  
%
Similarly in the scalar field model, 
we have shown that the ground state of $s$-wave superconductors is  from  $s$-wave, while the $p$-wave  and $d$-wave states  give excited states. 
We will also show that the traditional formalism of holographic $p$-wave superconductor is rather twisted in the gap structure so that one can not introduce the order parameter of $p_{x}\pm ip_{y}$ type. We will show how to fix this problem. 

Our investigation is done by constructing   the angle dependent gap function   in momentum space. 
For the vector field model, we will be able to construct the gap functions analytically by imposing the vortex free condition.   
For  the scalar field model case,   we can find general   $l$-wave type superconductors as excited states.
In general, the gap equation in holography is non-linearly coupled with that of the photon field in the AdS, which is not solvable by usual separation of variable, which is useful in linear equations. 
To overcome this difficulty, we consider the system near  the critical temperature ($T_c$) and expand the field and equation of motion by the small parameter $1-T/T_{c}$ and consider the system of equations order by order. 

We have also investigated $ T_{c }$ itself in the probe  limit of the gravity field    for  each models.  We have used matrix-eigenvalue algorithm with the Pincherle’s Theorem to calculate the $T_c$ values for scalar field model. For vector field model, we have used Sturm-Liouville eigenvalue method. Interestingly, {\it we have observed that the critical temperature for vector field model is higher than the critical temperature  of scalar field model} for any fixed value of the dimension of the Cooper pair operator.  We considered the possibility that $p$-wave condensation can be discussed  in scalar field model context provided s-wave condensation is forbidden under a special constraint due to e.g. the lattice symmetry.  
We   compared the gap functions for $p$-wave states coming from scalar  model and that from the vector model.   

This paper is organized as follows. We have started our discussion on different holographic set up for $s, p, d$-wave superconductors of the BCS theory in section 2. In section 3, we studied field equations of various holographic models and described their equation of motion in the unified fashion. 
In section 4, 
we have studied the vector field model with angular dependence  and show that the normalizable solution is available only  from the $p$-wave solution. 
We also show how to  introduce the $p_{x}\pm i p_{y}$ type of gap   of $p$-wave superconductivity in holographic set up. 
In section 5,  we compare the critical temperature for ground state of scalar field model and vector field model. We then compare the excited $p$-wave state from scalar field model with $p$-wave state in the vector field model.  We summarize our findings in section 6.

\section {Momentum dependence of the Gap in BCS theory} 
To understand the origin of the momentum dependence gap structure in $p$ and $d$-wave superconductors, we start with a basic discussion on the BCS gap structure. For arbitrary pairing interaction, the BCS gap equation can be written as \cite{anderson}
\begin{eqnarray}
	\Delta_k = -\frac{1}{2} \sum_{k'} \frac{V_{kk'}\Delta_{k'}}{\sqrt{\xi^2_{k'} + |\Delta_{k'}|^{2}}} \tanh\left(\frac{\sqrt{\xi^2_{k'} + |\Delta_{k'}|^{2}}}{2k_B T}\right)
	\label{eq28}
\end{eqnarray}
where $\xi_k$ is energy spectrum.
The potential in $k$-space can be expressed as
\begin{eqnarray}
	V_{kk'} =  
 \sum_{l=0}^{\infty}\frac{2l+1}{2}V_l(k,k')P_l(\hat{k}.\hat{k'}) ~~;~~ V_l(k,k')= \frac{4\pi}{\sqrt{kk'}}\int J_{l+\frac{1}{2}}(k'r) V(r) J_{l+\frac{1}{2}}(kr)r dr ~.
	\label{eq29}
\end{eqnarray}
where $l$ is angular quantum number and $V(r), P_l, J_{l}$ are the interaction potential, Legendre polynomials and Bessel function respectively. 
The different value of $l$ describes the different orbital symmetry which determines the different type superconductors, namely, $s$-wave($l=0$), $p$-wave($l=1$), $d$-wave($l=2$) superconductor. Recently $f$-wave($l=3$) and $g$-wave($l=4$) superconductors also has been reported in \cite{fwave1,fwave2} and \cite{gwave} respectively.
For non-zero value of $l$,  they lead to the momentum dependent BCS gap function.  For different orbital symmetry, the angle dependent gap structures are known in literature. The cuprate exhibits $d_{x^2-y^2}$-wave superconductivity.  The superconducting order parameter of $d_{x^2-y^2}$-wave superconductivity is \cite{dwave}
\begin{eqnarray}
	\Delta_{k} = \Delta(T) (\cos k_x  -\cos k_y ) \simeq \Delta(T) \cos(2\theta)~.
	\label{eq3}
\end{eqnarray}
 
This gap structure in the holographic set-up will play a very important role to understand the properties of real-world materials since there are many data available: those of  angle-resolved photoelectron spectroscopy (ARPES), Raman Spectroscopy, scanning tunnel spectroscopy and neutron magnetic scattering \cite{dwave}.   \\


\section{ Unified Field equations of holographic models}
We start with a general discussion from and summarizing    field equations   for   various spin fields in  a unified form 
 following  \cite{ghsc2}-\cite{ghsc3}.  We use planar symmetric AdS$_4$-Schwarzschild blackhole  as the background: 
\begin{eqnarray}
ds^2= -f(r)dt^2+ \frac{dr^2}{f(r)} + r^2 (dx^2 + dy^2) ~~, ~~~f(r)= r^2(1-\frac{r_h^2}{r^2})
\label{plnrm}
\end{eqnarray}
where $r_h$ is the horizon radius. For $s$-wave   superconductor,  we use  Abelian-Higgs model with a complex scalar field \cite{hs6a}:
\begin{eqnarray}
	\mathcal{L}_s= -\frac{1}{4} F^{\mu \nu} F_{\mu \nu} - (D_{\mu}\psi)^{*} D^{\mu}\psi-m^2 \psi^{*}\psi ,
	\label{ac0}
\end{eqnarray}
where $F_{\mu \nu}=\partial_{\mu}A_{\nu}-\partial_{\nu}A_{\mu}$, $D_{\mu}\psi=\partial_{\mu}\psi-iqA_{\mu}\psi$. From this Lagrangian,  the  field equations read
\begin{eqnarray}
	\partial_{\mu}[\sqrt{-g}F^{\mu\nu}] &=& 2\sqrt{-g} q^2 A^{\nu} \psi^{*} \psi + i q \sqrt{-g} [\psi^{*}\partial^{\nu}\psi -\psi \partial^{\nu}\psi^{*}] \nonumber \\
	\partial_{\mu}[\sqrt{-g}\partial^{\mu}\psi] &=& \sqrt{-g} [q^2 A_{\mu}A^{\mu} + m^2]\psi + iq [\sqrt{-g}(\partial^{\mu}\psi) A_\mu + \partial_{\mu}(\sqrt{-g}A^{\mu}\psi)]. 
	\label{ac1}
\end{eqnarray}
Using the ansatz $\psi=\psi(r)$ and $A=A_t(r)$, the equation motion for $s$-wave holographic superconductors takes form
\begin{eqnarray}
A^{\prime\prime}_t(r) + \frac{2}{r}A^{\prime}_t(r)- \frac{2q^2 \psi^*(r)\psi(r)}{f(r)}A_t(r) &=& 0 \\
\psi^{\prime\prime}(r) + \left[ \frac{f'(r)}{f(r)}+ \frac{2}{r}\right]\psi^{\prime}(r) + \left[\frac{q^2A^2_t(r)}{f^2(r)}- \frac{m^2}{f(r)}\right]\psi(r) &=& 0 ~.
\label{hscfe0}
\end{eqnarray}
For $p$-wave superconductors, the holographic $p$-wave model was first introduced in \cite{gubserp} using a $SU(2)$ Yang-Mills field in AdS$_4$-Schwarzschild background  with Lagrangian  
\begin{eqnarray}
\mathcal{L}_p^{YM} = -\frac{1}{4} F^{a}_{\mu\nu}F^{a\mu\nu}
\end{eqnarray}
where $F^a_{\mu\nu}=\partial_{\mu}A^a_{\nu}-\partial_{\nu}A^{a}_{\mu}+q\epsilon^{abc}A^{b}_{\mu}A^{c}_{\nu}$ is the field strength for $SU(2)$ gauge field.
The equation of motion yields
\begin{eqnarray}
\frac{1}{\sqrt{-g}}\partial_{\mu}[\sqrt{-g}F^{a\mu\nu}] + q \epsilon^{abc}A^{b}_{\mu}F^{c\mu\nu}=0 ~.
\end{eqnarray}
 To break the rotation symmetry in the system, the field ansatz is considered as $A=A_t(r)\sigma^3 dt + \psi^{(g)}_{x}(r)\sigma^1dx$ in which the condensed phase breaks $U(1)$ symmetry and $SO(2)$ rotational symmetry in $xy$-plane. Using this ansatz, the equation motion for non-abelian gauge theory reads
\begin{eqnarray}
A^{\prime\prime}_t(r) + \frac{2}{r}A^{\prime}_t(r)- \frac{q^2 \psi^{(g)*}_x(r)\psi^{(g)}_x(r)}{r^2 f(r)}A_t(r) &=& 0 \\
\psi^{(g)\prime\prime}_x(r) + \frac{f'(r)}{f(r)}\psi^{(g)\prime}_x(r) + \frac{q^2A^2_t(r)}{f^2(r)}\psi^{(g)}_x(r) &=& 0 ~.
\end{eqnarray}
Later, an alternative $p$-wave holographic superconductors model was introduced by a charged vector field. The matter Lagrangian density for this model is 
\begin{eqnarray}
\mathcal{L}^V_{p}= -\frac{1}{4}F_{\mu\nu}F^{\mu\nu}-\frac{1}{2}\Psi^{\dagger}_{\mu\nu}\Psi^{\mu\nu}- m^2 \psi^{\dagger}_{\mu}\psi^{\mu}
\label{pwavev0}
\end{eqnarray}
 which gives the following field equations
\begin{eqnarray}
\label{pwavev1}
\frac{1}{\sqrt{-g}}\partial_{\mu}[\sqrt{-g}F^{\mu\nu}]+ i[\psi^{\dagger}_{\mu} \Psi^{\mu\nu}- \psi_{\mu}(\Psi^{\mu\nu})^{\dagger}] &=& 0 \\
\frac{1}{\sqrt{-g}}\partial_{\mu}[\sqrt{-g}\Psi^{\mu\nu}]- [m^2 \psi^{\nu}+i A_{\mu} \Psi^{\mu\nu}] &=& 0
\label{pwavev2}
\end{eqnarray}
where $\Psi_{\mu\nu}=\partial_{\mu}\psi_{\nu}-\partial_{\nu}\psi_{\mu}-i A_{\mu}\psi_{\nu}+i A_{\nu}\psi_{\mu}$. By taking complex vector field $\psi_{\mu}dx^{\mu}=\psi_x (r)dx^{x}$ along $x$-direction, we break the rotational symmetry of the vector field. Using this vector field ansatz and gauge field ansatz $A_{\mu}dx^{\mu}=A_t(r)dt$, the field equations takes form 
\begin{eqnarray}
A^{\prime\prime}_t(r) + \frac{2}{r}A^{\prime}_t(r)- \frac{2 q^2 \psi^*_x(r)\psi_x(r)}{r^2 f(r)}A_t(r) &=& 0 \\
\psi^{\prime\prime}_x(r) + \frac{f'(r)}{f(r)}\psi^{\prime}_x(r) + \left[\frac{q^2A^2_t(r)}{f^2(r)}-\frac{m^2}{f(r)}\right]\psi_x(r) &=& 0 ~.
\end{eqnarray}
If we map $\psi_x(r)=\frac{\psi^{(g)}_x(r)}{\sqrt{2}}$, we will recover the field equations for non-abelian model. This two models are equivalent for $m^2=0$. To construct holographic $d$-wave model, the vector field model was generalized tensor field model with minimal effective matter Lagrangian density
\begin{eqnarray}
\mathcal{L}^T_d=-\frac{1}{4}F_{\mu\nu}F^{\mu\nu}-(D_{\mu}B_{\nu\lambda})^{\dagger}D^{\mu}B^{\nu\lambda}- m^2 B^{\dagger}_{\mu\nu}B^{\mu\nu}
\end{eqnarray}  
where $B^{\mu\nu}$ is a charged tensor field. To realized $d$-wave condensate, they considered the tensor field ansatz $B_{xx}=-B_{yy}=\psi_{xx}(r)$ which breaks rotational symmetry and flips sign under a $\frac{\pi}{2}$-rotation on the $xy$-plane. Using this tensor field ansatz and the gauge field ansatz $A=A_t(r)$, the field equations read 
\begin{eqnarray}
A^{\prime\prime}_t(r) + \frac{2}{r}A^{\prime}_t(r)- \frac{4q^2 \psi^*_{xx}(r)\psi_{xx}(r)}{r^4 f(r)}A_t(r) &=& 0 \\
\psi^{\prime\prime}_{xx}(r) + \left[ \frac{f'(r)}{f(r)}- \frac{2}{r}\right]\psi^{\prime}_{xx}(r) + \left[\frac{q^2A^2_t(r)}{f^2(r)}- \frac{m^2}{f(r)}-\frac{2f^{\prime}(r)}{rf(r)}\right]\psi_{xx}(r) &=& 0 ~.
\label{dhsctm}
\end{eqnarray}
This model is based on minimal effective action without looking the constraint equations for propagating degrees of freedom. Another holographic $d$-wave tensor field model was proposed with the correct number of propagating degrees of freedom in \cite{dhsc12} \footnote{ The Lagrangian density in this modified model is
$
\mathcal{L}_d^{H} = -\frac{1}{4}F_{\mu\nu}F^{\mu\nu}- |D_{\alpha}B_{\mu\nu}|^2+2 |D_{\mu}B^{\mu\nu}|^2+ |D_{\mu}B|^2-[D_{\mu}B^{\dagger\mu\nu}D_{\nu}B+ h.c.] 
 -iq F_{\mu\nu}B^{\dagger\mu\lambda}B^{\nu}_{\lambda} -m^2(|B_{\mu\nu}|^2 - |B|^2)+ 2 R_{\mu\nu\rho\lambda}B^{\dagger\mu\rho}B^{\nu\lambda}- \frac{1}{4} R |B|^2
$
where $B \equiv B^{\mu}_{~\mu}, ~B_{\mu}\equiv D^{\nu}B_{\nu\mu}$ and $R_{\mu\nu\rho\lambda}$ is the Riemann tensor of the background spacetime.
With the tensor fields ansatz $B_{xy}=\psi_{xy}(r)$ and gauge field ansatz $A=A_t(r)$, the matter field equation reads
$$
\psi^{\prime\prime}_{xy}(r) + \left[ \frac{f'(r)}{f(r)}- \frac{2}{r}\right]\psi^{\prime}_{xy}(r) + \left[\frac{q^2A^2_t(r)}{f^2(r)}- \frac{m^2}{f(r)}-\frac{2f^{\prime}(r)}{rf(r)}+\frac{2}{r^2}\right]\psi_{xy}(r) = 0 
\label{dhschm}
$$
which is differ from the eq.(\ref{dhsctm}) because of the last term. If we take $\psi_{xy}(r)=\frac{r^2}{\sqrt{2}}\psi(r)$, we will get exactly same field equations for $s$-wave holographic model.}. To generalized spin field models with spin $s$, we consider the field eq.(\ref{dhsctm}) for $d$-wave holographic superconductors model. 
Using the mapping 
\begin{eqnarray}
\psi_s(r)=\psi(r), ~~~\psi_s(r)=\frac{\psi_x(r)}{r} ~~~\text{and}~~\psi_s(r)=\frac{\sqrt{2}\psi_{xx}(r)}{r^2}
\end{eqnarray}
the unified form of the spin fields equation for holographic superconductor models with different wave state ($s, p, d$-wave respectively) takes in the following form \cite{ghsc2}
\begin{eqnarray}
	\label{spin1}
	A_t^{\prime\prime}(r) + \frac{2}{r}A^{\prime}_t(r)- \frac{2q^2 \psi^*_s(r)\psi_s(r)}{f(r)}A_t(r)=0 \\
	\psi^{\prime\prime}_s(r) + \left[ \frac{f'(r)}{f(r)}+ \frac{2}{r}\right]\psi^{\prime}_s(r) + \left[\frac{q^2A^2_t(r)}{f^2(r)}- V_s(r)\right]\psi_s(r) =0
	\label{spin2}
\end{eqnarray}
where 
\begin{eqnarray}
	V_s(r)= \frac{m^2}{f(r)} - s(2-s)\frac{f'(r)}{rf(r)}+ \frac{s(s-1)}{r^2}
	\label{spin3}
\end{eqnarray}
which is called `effective potential' \footnote{This terminology is quoted because it is not exactly the same as the actual effective potential term derived from a dynamical equation. For our discussion, we consider it as an effective potential for this coupled equation.}. From this above unified field equation, we recover the field equations for $s, p, d$-wave holographic superconductor for the spin values $s=0, 1, 2$ respectively.
If we calculate the perturbation of Maxwell's field $A_{\mu}= \delta^x_{\mu}A_{x}\exp(-i\omega t)$ for conductivity, we will get the same equation structure for different values of spin. The radial (only) dependent field structure in the unified field equations leads to spherically symmetric ground state of superconductors. Since all holographic superconductor models  so far, are governed by the above unified field equations which depends only on $r$, we can say that the momentum dependent order parameter in holographic set-up is missing in the literature.  

To gain a better understanding of high $T_c$ superconductors through holographic set-up, we need to modify the ansatz of the fields which will help us to distinguish $s, p, d$-wave superconductors in a generic sense. The generic sense means that the distinguishable properties for $s, p, d$-wave superconductors depends on the values of angular momentum quantum number instead of spin number $s$ in holographic set-up. The angular momentum number determines the actual orbital symmetry which is responsible for the superconductivity.  

Although the choice of the ansatz of the field breaks the rotational symmetry in holographic superconductors models, those models do not have any angular momentum number $l$ which is essential to understand the $p$-wave or $d$-wave superconductivity. In \cite{dhsc1}, the spatial angle $\theta$ is introduced by the transformation of spin two field and the angle dependence gap is generated using the interaction term between the spin two field with fermions. From the literature of holographic superconductors, it seems that the spin number $s$ of the field in holographic set-up is related to the angular momentum number $l$ of the boundary theory. In order to understand the connection between them, we start with vector field ($s=1$) model with angular dependent fields.

\section{Vector field model with angle dependent Gap}
\noindent 
Here, we assume that there is strong asymmetry in the direction of the c-axis so that we can just consider 2+1 dimensional direction. 
To introduce the angle dependent gap structure in holographic superconductors, we consider the polar coordinate of $2+1$-dimensional boundary where  the system  lives.
Accordingly we write the AdS$_4$-Schwarzschild black hole metric (\ref{plnrm}) in polar coordinate $(t,r, u, \theta)$ reads
\begin{eqnarray}
	ds^2=-f(r)dt^2+\frac{1}{f(r)}dr^2+ r^2 (du^2 + u^2 d\theta^2)  ~~;~~ f(r)=r^2\left( 1- \frac{r^3_{h}}{r^3} \right) ,	\label{m10}
\end{eqnarray}
%
%
from which the Hawking temperature can be read as 
\begin{eqnarray}
	T_{H} = \frac{f^{\prime}(r_{h})}{4\pi}=\frac{3r_{h}}{4\pi}~.
	\label{gzx1}
\end{eqnarray} 
The $p$-wave holographic superconductors model has been described by the matter Lagrangian density (\ref{pwavev0}) which consists of gauge field and vector field. The field equation are given by eq.(\ref{pwavev1}) and eq.(\ref{pwavev2}). We now modify the complex vector field and gauge field ansatz which reads
$$\psi_{\mu}=\psi_x(r,x,y)dx, ~~~\text{and}~~~ A_{\mu}=A_t(r,x,y)dt ~.$$
\noindent We consider field along one direction since we want to break the rotational symmetry of the vector field. 
The ansatz changes in polar coordinate 
\begin{eqnarray}
	A_{\mu}&=&A_t(r, x, y)dt=A_t(r, u, \theta)dt \\
	\psi_{\mu}&=&\psi_x(r, x, y)dx=\psi_x\{\cos\theta du - u\sin\theta d\theta\}= \psi_u(r, u, \theta) du + \psi_{\theta}(r, u, \theta)d\theta 
\end{eqnarray}
where 
\begin{eqnarray}
	\psi_u(r, u, \theta) = \cos\theta \psi_x(r, u, \theta) ~ \text{and}~~ \psi_{\theta}(r, u, \theta) = -u\sin\theta \psi_{x}(r, u, \theta) ~.
	\label{eq70}
\end{eqnarray}
Using the above relation, we find the relation between $\psi_u$ and $\psi_{\theta}$ which is
\begin{eqnarray}
	\psi_{\theta}(r, u, \theta) = -u \tan\theta \psi_u(r, u, \theta) ~.
	\label{relutheta}
\end{eqnarray}
Using the eq.(\ref{pwavev1}) and the above field ansatz, the gauge field equation becomes
\begin{eqnarray}
	\partial^2_{r} A_t + \frac{2}{r} \partial_{r}A_t +\frac{1}{r^2 f(r)}\left[\partial^2_{u}A_t + \frac{\partial_{u}A_t}{u}+ \frac{\partial_{\theta}^2A_t}{u^2}\right] = \frac{2q^2}{u^2r^2 f(r)}\left[u^2 |\psi_u|^2 +|\psi_{\theta}|^2\right]A_t 
	\label{eq072}
\end{eqnarray}
From eq.(\ref{pwavev2}), the matter field equation reads
\begin{eqnarray}
	\frac{1}{\sqrt{-g}}\partial_{\mu}[\sqrt{-g}\Psi^{\mu\nu}]- [m^2 \psi^{\nu}+iq A_{\mu} \Psi^{\mu\nu}] &=& 0
\end{eqnarray}
Setting $\nu=u$ and substituting $\sqrt{-g}=r^2u$, we obtain
\begin{eqnarray}
	\partial^2_r \psi_u+ \frac{f'(r)}{f(r)}\partial_r \psi_u +\frac{1}{r^2f(r) u^2 }\partial_{\theta}[\partial_{\theta}\psi_u -\partial_{u}\psi_{\theta}] + \left[\frac{q^2A^2_{t}}{f^2(r)}-\frac{m^2}{f(r)}\right]\psi_{u} = 0 &&~.
	\label{eq73}
\end{eqnarray}
Similarly we obtain field equation for $\psi_{\theta}$ by setting $\nu=\theta$, 
\begin{eqnarray}
	\partial^2_r \psi_{\theta}+ \frac{f'(r)}{f(r)}\partial_r \psi_{\theta} +\frac{u}{r^2f(r)}\partial_{u}\left(\frac{\partial_{u}\psi_{\theta}-\partial_{\theta}\psi_{u}}{u}\right)+ \left[\frac{q^2A^2_{t}}{f^2(r)}-\frac{m^2}{f(r)}\right]\psi_{\theta} = 0 &&
	\label{eq74}
\end{eqnarray}
\subsection{The angular dependent part of the matter field}
We now impose vortex free condition (perpendicular to the plane) which is
\begin{eqnarray}
	\partial_{u}\psi_{\theta}(r, u, \theta)-\partial_{\theta}\psi_{u}(r, u, \theta)=0~.
	\label{vortexfree0}
\end{eqnarray} 
Using the relation between $\psi_{u}$ and $\psi_{\theta}$, the above condition becomes
\begin{eqnarray}
	u \partial_{u}\psi_u(r,u,\theta)+ \psi_u(r,u,\theta) +\frac{\partial_{\theta}\psi_{u}(r,u,\theta)}{\tan\theta}=0
	\label{vortexfree1}
\end{eqnarray}
Using this condition, we want to solve the $u,\theta$ dependence part of the matter field. We can write
\begin{eqnarray}
	\psi_u(r,u,\theta)=\Psi(r)\mathcal{R}_u(u,\theta)=\Psi(r) U(u)\Theta(\theta) ~.
	\label{vortexfree3}
\end{eqnarray} 
From eq.(\ref{vortexfree1}) and eq.(\ref{vortexfree3}), we now try to solve the boundary wave state $\mathcal{R}_u(u, \theta)$ with help of the separation constant $l_p$ 
\begin{eqnarray}
	\frac{1}{\Theta(\theta)}\frac{\partial_{\theta}\Theta(\theta)}{\tan\theta}=-l_p , ~~~~~~~~~~~~~~~&;&~~~~~~~~~~u \partial_{u}U(u) +(1-l_p) U(u)=0, \\
	\Rightarrow \Theta(\theta)= (\cos\theta)^{l_p} ,~~~~~~~~~~~~~~~&;&~~~~~~~~\Rightarrow U(u)= u^{l_p-1}. 
\end{eqnarray} 
 Therefore the solution   reads
$	\mathcal{R}_u(u,\theta)=   u^{l_p-1} (\cos\theta)^{l_p}.
$
where  we deleted one multiplicative  integration constants absorbing them into $\Psi$. Using the relation (\ref{relutheta}) and $\psi_{\theta}(r, u, \theta)=\Psi(r)\mathcal{R}_{\theta}(u,\theta)$, we   find $\mathcal{R}_{\theta}(u,\theta)=   u^{l_p} (\cos\theta)^{l_p-1} \sin\theta$.
Therefore, we can write
\begin{eqnarray}
	\psi_{u}(r,u,\theta)= \Psi(r)  u^{l_p-1} (\cos\theta)^{l_p}~~,~~~~\psi_{\theta}(r,u,\theta)=C\Psi(r) u^{l_p} (\cos\theta)^{l_p-1} \sin\theta.
	\label{eq81}
\end{eqnarray}
The separation constant $l_p$ should be integer: this can be seen from the fact that the matter field  should be one valued under the rotation of $2\pi$ of $\theta$, which is the same as  the twice of the $\pi$ rotation under which 
\be 
\cos\theta \to -\cos\theta, 
\ee
 From this, we can identify   $l_p$ as the angular momentum in this   set-up. For   $l_p=0$, 
\begin{eqnarray}
	\psi_{u}(r,u,\theta)= \frac{\Psi(r)}{u} ~~~~\text{and}~~~~ 	\psi_{\theta}(r,u,\theta)= C\Psi(r)\tan\theta
\end{eqnarray} 
which is not normalizable because the normalization condition is  
\be 
\int \psi^2_x(r,x,y) dx dy= \int_{0} \int_{0} \left[\psi^2_u(r, u, \theta)+\frac{\psi^2_{\theta}(r, u, \theta)}{u^{2}}\right]udu d\theta =\int_{0}^{a} \frac{du}{u}   |\Phi(r)|^{2} 
 ,\ee
 which is   logarithmically divergent. 
 Therefore the vector field model does not give us the normalizable ground state for the s-wave state. 
For  the  $p$-wave of $l_{p}=1 $,    the solution is given by 
 \begin{eqnarray}
	\psi_{u}(r,u,\theta)=  \cos\theta \Psi(r),  ~~~~~~\psi_{\theta}(r,u,\theta)=  -u  \sin\theta\Psi(r), 
\end{eqnarray}
which is normalizable solution.   
Notice that $C$ should be chosen to be $-1$ to be consistent with  eq.(\ref{eq70}). 
Therefore the ground state of the    vector field model comes from  the $l_p=1$, while the s-wave solution of the model is not normalizable. 
  which is   first  main result of this paper.
 Notice also that  $\psi_{x}(r, u, \theta)=\Psi(r)$  for $l_p=1$ but only for this case. That is, the solution  $\psi_x(r, x, y) dx$ is reduced to  $\psi_x(r) dx$ for the ground state. 
 Now  if we take the $\psi_{x}(r)$ as the order parameter of the $p$-wave superconductivity as it was suggested in the original model \cite{gubserp}, there is no angular dependence in the gap,  which is a contradiction. 
 {  Our analysis in the present setup is telling us that the gap function of the $p$-wave model is $\psi_{u}$, not the $\psi_{x}$. } 
Notice that 
 $\psi_{x}$ can not be the order parameter of  p-wave superconductivity  because it does not have any angular dependence, and  $\psi_{\theta}$ can not be  the one either, because it shows the vanishing gap   at $u=0$, which is a coordinate singularity not the real nature. 
 
 There is nothing wrong here but what we got is not really what we would expect in the usual tensor analysis. 
 All the oddities come from the assumption that 
only $\psi_{x}\neq 0$ while $\psi_{i}=0, i=y,t,r$ which is very unusual gauge choice from rotation tensor point of view. 
The better    ansatz for the gap structure should be  the following one:   
    \be
    \psi_{x}= A(r)\cos\theta,\quad \psi_{y}=A(r)\sin\theta.
    \ee
Then by a simple calculation,  we  can get the identification  $A=\Psi(r)$ where $\Psi(r)$ is the function we met before.  Here we can regards any of
 $A_{i} $ as the order parameter. Then, the order parameter for the gap structure $p_{x}+ip_{y}$ can be naturally introduced as 
 \be
 \psi_{x}+i\psi_{y}=\Phi(r) e^{i\theta},
 \ee
  which has not been possible so far. This is  simple but one of the main points of this paper.

\subsection{The critical temperature and field solutions}
 We now proceed to solve the radial part of the matter field. 
\noindent Using the vortex free condition, eq.(s)(\ref{eq73},\ref{eq74}) can be written as
\begin{eqnarray}
	\label{eq76}
	\partial^2_r \psi_u(r, u,\theta) + \frac{f'(r)}{f(r)}\partial_r \psi_u(r, u,\theta) + \left[\frac{q^2A^2_{t}}{f^2(r)}-\frac{m^2}{f(r)}\right]\psi_{u}(r, u,\theta) &=& 0 \\
	\partial^2_r \psi_{\theta}(r, u,\theta)+ \frac{f'(r)}{f(r)}\partial_r \psi_{\theta}(r, u,\theta) + \left[\frac{q^2A^2_{t}}{f^2(r)}-\frac{m^2}{f(r)}\right]\psi_{\theta}(r, u,\theta) &=& 0
	\label{eq77}
\end{eqnarray}
Substitute eq.(\ref{eq81}) in eq.(\ref{eq76}) and eq.(\ref{eq77}), we get a single equation for $\Psi(r)$ which takes form as
\begin{eqnarray}
	\Psi^{\prime\prime}(r) +\frac{f'(r)}{f(r)} \Psi^{\prime} + \left[\frac{q^2A^{(0)2}_{t}(r)}{f^2(r)}-\frac{m^2}{f(r)}\right]\Psi(r) = 0
	\label{orpeq1}
\end{eqnarray}
where we have substitute the zeroth order of gauge field part from the gauge field expansion $A_t(r, u,\theta)=A^{(0)}_t (r) + \epsilon A^{(1)}_t(r, u,\theta) $ near the critical temperature. Substitute $\psi_u(r,u,\theta)$ and $\psi_{\theta}(r,u,\theta)$ for $l_p=1$, the zeroth order gauge field equation becomes
\begin{eqnarray}
	A_t^{(0)\prime\prime}(r) + \frac{2}{r} A_t^{(0)\prime}(r) = \frac{2q^2}{r^2 f(r)} |\Psi(r)|^2 A_t^{(0)}(r)
	\label{orpeq2}
\end{eqnarray}
The above two field equations are same with the field equations for vector field model in literature. Therefore, the critical temperature and the temperature dependence condensation operator value will be unchanged.  To get the solution of the fields, we need to know the asymptotic behavior of the fields.  At the asymptotic limit, we consider $f(r)\approx r^2 $ and $\mathcal{O}(r^4)=\infty$.
Using this, we obtain the field equation near boundary 
\begin{eqnarray}
	\Psi^{\prime\prime}(r) + \frac{2}{r} \Psi^{\prime}(r) -\frac{m^2}{r^2}\Psi(r) &=& 0 \\
	A_t^{(0)\prime\prime}(r) + \frac{2}{r} A_t^{(0)\prime}(r)  &=& 0
\end{eqnarray}
Using the gauge/gravity duality, the asymptotic behavior of the field reads
\begin{eqnarray}
	\Psi (r) = \frac{\Psi_{-}}{r^{\delta_{-}-1}} + \frac{\Psi_{+}}{r^{\delta_{+}-1}} ~~~~~~~;~~~~~~A^{(0)}_t(r) = \mu - \frac{\rho}{r}
	\label{asyp}
\end{eqnarray}
where $\delta_{\pm}=\frac{1}{2}[3\pm\sqrt{1+ 4m^2}]$ is the scaling dimension, $\mu$ and $\rho$ is the chemical potential and the charge density respectively and
 $\Psi_{\pm}$ maps to temperature dependent condensation operator value of the boundary theory. The Breitenlohner-Freedman mass bound \cite{bf1},\cite{bf2} for this holographic set-up is $m^2_{BF}\geq -\frac{1}{4} $. Under the coordinate transformation $z=\frac{r_h}{r}$, the metric field reads
 \begin{eqnarray}
 	f(z)=\frac{r_h^2}{z^2} g(z) ~~~~~;~~~~g(z)=1-z^3 ~.
 	\label{gz1}
 \end{eqnarray}
In the $z$-coordinate, the field equations become
\begin{eqnarray}
	\frac{d^2\Psi(z)}{dz^2} + \frac{g'(z)}{g(z)}\frac{d\Psi(z)}{dz} + \left[\frac{q^2 (A^{(0)}_t(z))^2}{r^2_h g^2(z)}-\frac{m^2}{z^2 g(z)}\right] \Psi(z) &=& 0 \\
	\frac{d^2A^{(0)}_t(z)}{dz^2} - \frac{2 q^2}{r^2_{h}g(z)} |\Psi(z)|^2 A_t^{(0)}(z) &=& 0
	\label{eq117}
\end{eqnarray}
At $T=T_c$, the matter field $\Psi(z)=0$ which leads to the zeroth order  gauge field eq.(\ref{eq117}) 
\begin{eqnarray}
	\frac{d^2A^{(0)}_t(z)}{dz^2} =0 ~.
\end{eqnarray}
Using the aymptotic behavior of fields (\ref{asyp}), the solution of the above equation reads
\begin{eqnarray}
	A^{(0)}_t(z) = \lambda r_{h(c)} (1-z) ~.
\end{eqnarray}
where $\lambda =\frac{\rho}{r^2_{h(c)}}$.
We can write the radial part of the  matter field in following form
\begin{eqnarray}
	\Psi(z)=\frac{\mathcal{C}_p}{r^{\delta-1}_h} z^{\delta-1} F(z)
	\label{sol1}
\end{eqnarray}
where $F(z)$ is the trail function for Sturm-Liouville eigenvalue method, $\delta (=\delta_{\pm})$ is the scaling dimension  and  $\mathcal{C}_p$ is unknown constant which need to be determined. We now substitute this in the matter field equation which yields
\begin{eqnarray}
	\frac{d^2F(z)}{dz^2} +  \left[\frac{2(\delta-1)}{z}+\frac{g^{\prime}(z)}{g(z)} \right]\frac{dF(z)}{dz} &+& \left[ \frac{(\delta-1)(\delta-2)}{z^2}+\frac{g^{\prime}(z)}{g(z)} \frac{(\delta-1)}{z}-  \frac{m^2 }{z^2g(z)}\right]F(z) \nonumber \\
	&+& \frac{\tilde{\lambda}^2(1-z)^2}{g^2(z)}F(z) =0 
\end{eqnarray} 
where $\tilde{\lambda} = q\lambda$.
The above equation can be written in the Sturm-Liouville form 
\begin{eqnarray}
	\frac{d}{dz}\left\{p(z)F'(z)\right\}+q(z)F(z)+\tilde{\lambda}^2 r(z)F(z)=0
	\label{sturm}
\end{eqnarray}
with 
\begin{eqnarray}
	p(z)&=&z^{2\delta-2}g(z)  ~,~~~~~~~~~r(z)=\frac{z^{2\delta-2}}{g(z)} (1-z)^2\nonumber\\
	q(z)&=&z^{2\delta-2}g(z)\left\{ \frac{(\delta-1)(\delta -2)}{z^2} +\frac{g'(z)}{g(z)}\frac{(\delta-1)}{z}-\frac{m^2}{g(z) z^2} \right\} ~.
	\label{i1}
\end{eqnarray}
The above identification enables us to write down an equation for the eigenvalue $\tilde{\lambda}^2$ which minimizes the expression 
\begin{eqnarray}
	\tilde{\lambda}^2 &=& \frac{\int_0^1 dz\ \{p(z)[F'(z)]^2 - q(z)[F(z)]^2 \} }
	{\int_0^1 dz \ r(z)[F(z)]^2}~.
	\label{eq5abc}
\end{eqnarray}
For the estimation of $\tilde{\lambda}^{2}$, we shall now use the trial function $F= F_{\tilde\alpha} (z) \equiv 1 - \tilde\alpha z^2$ which   satisfies the conditions $F(0)=1$ and $F'(0)=0$.
The critical temperature reads from eq.(\ref{gzx1})
\begin{eqnarray}
	T_c = \frac{3}{4\pi} \frac{\sqrt{q \rho}}{\sqrt{\tilde{\lambda}}}
	\label{new21}
\end{eqnarray}
where the Sturm-Liouville eigenvalue $\tilde{\lambda}$ is estimated from eq.(\ref{eq5abc}). For $m^2=-\frac{3}{16}$, the scaling dimensions are $\delta=\delta_{-}=\frac{5}{4}$ and $\delta=\delta_{+}=\frac{7}{4}$. We have shown the critical temperatures in the Table \ref{tab1}. The critical temperature for $\delta=2 ~(\text{setting}~ m^2=0)$ is $T_c=0.124 \sqrt{\rho}$ which matches with the result from the non-abelian model \cite{gubserp}. \\
\begin{table}[h!]
	\centering
	\begin{tabular}{| c | c | c | c |}
		\hline
		$\delta$ & $\tilde{\lambda}^2$ & $\tilde{\alpha}$ & $T_c$\\
		\hline 
		$\frac{5}{4}$ & 1.2418 & 0.1426 & $T_c =0.226 \sqrt{q\rho}$\\
		\hline
		$\frac{7}{4}$ & 7.8766 & 0.4017 & $T_c =0.142 \sqrt{q\rho}$ \\
		\hline
	\end{tabular}
	\caption{Critical temperature for different scaling dimension for $m^2=-\frac{3}{16}$.}
	\label{tab1}
\end{table}

\noindent We now move to calculate the constant $\mathcal{C}_p$ from the gauge field equation near $T_c$. Substituting eq.(\ref{sol1}) in eq.(\ref{eq117}), we get
\begin{eqnarray}
	\frac{d^2\mathcal{A}^{(0)}_t(z)}{dz^2}  &=& \frac{\mathcal{C}^{2}_p}{r^{2\delta}_{h}} \mathcal{B}(z) \mathcal{A}^{(0)}_t(z)
	\label{cx1} 
\end{eqnarray}
where $\mathcal{B}(z)= 2q^2 z^{2\delta-2}\frac{F^{2}(z)}{g(z)}$. We may now expand $\mathcal{A}^{(0)}_t(z)$ in the small parameter $\frac{\mathcal{C}^{2}_p}{r^{2\delta}_{h}}$ as 
\begin{eqnarray}
	\frac{\mathcal{A}^{(0)}_t(z)}{r_{h}} = \lambda  (1-z)  + \frac{\mathcal{C}^{2}_p}{r^{2\delta}_{h}} \chi (z)
	\label{cx2} 
\end{eqnarray}
with $\chi (1)= 0 =\chi^{\prime}(1)$. 
From eq.(\ref{cx2}), we get the asymptotic behavior (near $z=0$) of the gauge field. Comparing the both equations of the gauge field about $z=0$, we obtain 
\begin{eqnarray}
	\mu -\frac{\rho}{r_{h}}z &=& \lambda r_{h} (1-z)  + \frac{\mathcal{C}^{2}_p}{r^{2\delta-1}_{h}} \left\{\chi(0)+z\chi^{\prime}(0)+... \right\}
	\label{cx7}
\end{eqnarray}
Comparing the coefficient of $z$ on both sides of eq.(\ref{cx7}), we obtain
\begin{eqnarray}
	-\frac{\rho}{r^2_{h}} = -\lambda  + \frac{\mathcal{C}^{2}_p}{r^{2\delta}_{h}}\chi_s^{\prime}(0) ~.
	\label{cx8}
\end{eqnarray}
We now need to find out the $\chi^{\prime}(0)$ by substituting eq.(\ref{cx2}) in eq.(\ref{cx1}).
Comparing the coefficient of $\frac{\mathcal{C}^{2}_p}{r^{2\delta}_{h}}$ of left hand side and right hand side of the eq.(\ref{cx1}), we get the equation for the correction $\chi(z)$ near to the critical temperature
\begin{eqnarray}
	\chi^{\prime \prime}(z)  = \lambda \mathcal{B}(z)(1-z) ~.
	\label{cx3}
\end{eqnarray}
Using the boundary condition of $\chi(z)$, we integrate (\ref{cx3}) between the limits $z=0$ and $z=1$ which gives
\begin{eqnarray}
	\chi^{\prime}(z)\mid_{z\rightarrow 0} = -\lambda  \mathcal{B}_{s}
	\label{cx5}
\end{eqnarray}
where $\mathcal{B}_{s} = 2q^2 \int^{1}_{0} dz \frac{ z^{2\delta-2} (1-\tilde{\alpha}z^2)^2}{1+z+z^2} $. 
Using eq.(\ref{cx5}) and eq.(\ref{cx8}), we obtain 
\begin{eqnarray}
	\mathcal{C}^{2}_p &=& \frac{r_h^{2\delta}}{\mathcal{B}_s}\left[\frac{r^2_{h(c)}}{r^2_h}-1\right]
	\label{cx9}
\end{eqnarray}  
where the definition of $\lambda$ is used. Using the expression for the critical temperature eq.(\ref{cx9}), we get
\begin{eqnarray}
	\mathcal{C}^{2}_p = \frac{(4\pi T)^{2\delta}}{\mathcal{B}_{s}[3]^{2\delta}}\left(\frac{T_{c}}{T}\right)^{2} \left[1- \left(\frac{T}{T_{c}}\right)^{2} \right] ~.
	\label{cx10}
\end{eqnarray}
Using the fact that $T \approx T_{c}$, we can write
$	T^{2\delta}\left(\frac{T_{c}}{T}\right)^{2} \left[1- \left(\frac{T}{T_{c}}\right)^{2} \right] \approx 2 T_c^{2\delta}\left[1- \left(\frac{T}{T_{c}}\right)\right] ~.$
Using this, we finally obtain the constant which gives the temperature dependence condensation value in following form
\begin{eqnarray}
	\mathcal{C}_p  = \sqrt{\frac{2}{\mathcal{B}_s}} \left[\frac{4\pi}{3}\right]^{\delta} T^{\delta}_{c} \sqrt{1-\frac{T}{T_{c}}}= \beta T^{\delta}_{c} \sqrt{1-\frac{T}{T_{c}}}~.
	\label{cx12}
\end{eqnarray}
where $\beta= \sqrt{\frac{2}{\mathcal{B}_s}} \left[\frac{4\pi}{3}\right]^{\delta}$. Near the critical temperature, the radial part of the scalar field solution now takes form 
\begin{eqnarray}
	\Psi(z)= \tilde{\beta} T_c\sqrt{1-\frac{T}{T_{c}}} z^{\delta-1} (1- \tilde{\alpha} z^2)= \tilde{\beta} T_c\sqrt{1-\frac{T}{T_{c}}}  \tilde{F}(z)
	\label{eq.4.49}
\end{eqnarray}
where $\tilde{\beta}=\sqrt{\frac{2}{\mathcal{B}_s}} \left[\frac{4\pi}{3}\right]$ and $\tilde{F}(z)=z^{\delta-1} (1- \tilde{\alpha} z^2)$. Given value of $m^2$ and $\delta$, the value of $\tilde{\beta}$ and $\tilde{\alpha}$ are fixed from the SL method. For $m^2=-\frac{3}{16}$, we get the value of $\tilde{\beta}= 7.359$ and $\tilde{\beta}=11.96$ for $\delta_{-}=\frac{5}{4}$ and $\delta_{+}=\frac{7}{4}$ respectively.  \\

\subsection{Gap structure in vector field model}
\begin{figure}[h!]
	\centering
	\begin{subfigure}[b]{0.3\textwidth}
		\centering
		\includegraphics[scale=0.4]{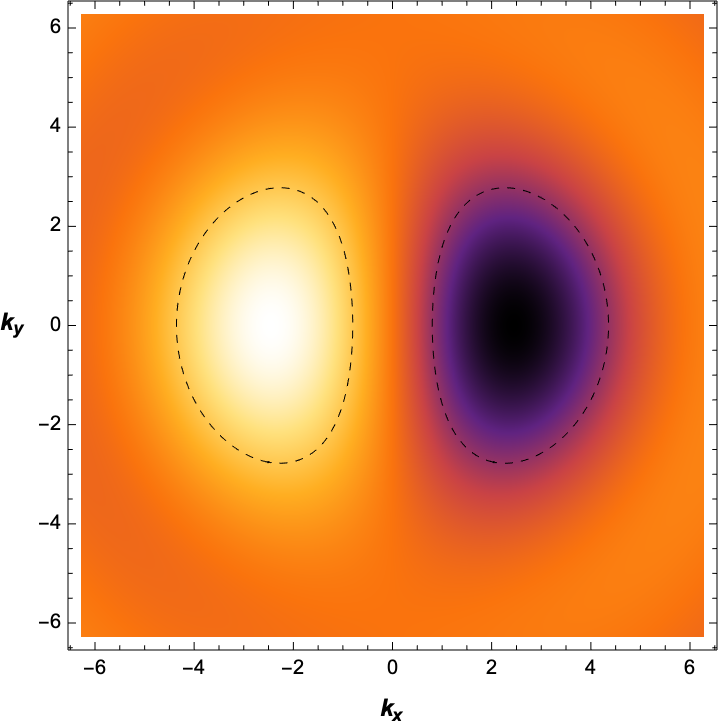}
		\caption{$\Delta^{(u)}_{k}$}
		\label{a1111pv}
	\end{subfigure}
	\hfil
	\begin{subfigure}[b]{0.3\textwidth}
		\centering
		\includegraphics[scale=0.4]{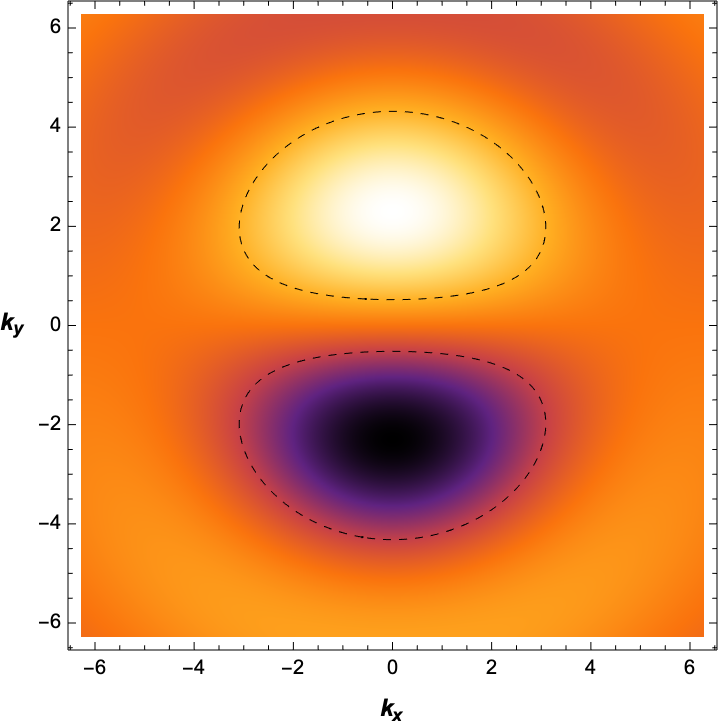}
		\caption{$\Delta^{(\theta)}_{k}$}
		\label{a2111pv}
	\end{subfigure}
	\caption{Two components gap function from vector wave model with $l_p=1$.}
	\label{fig15pv}
\end{figure}

\begin{figure}[h!]
	\centering
	\begin{subfigure}[b]{0.3\textwidth}
		\centering
		\includegraphics[scale=0.45]{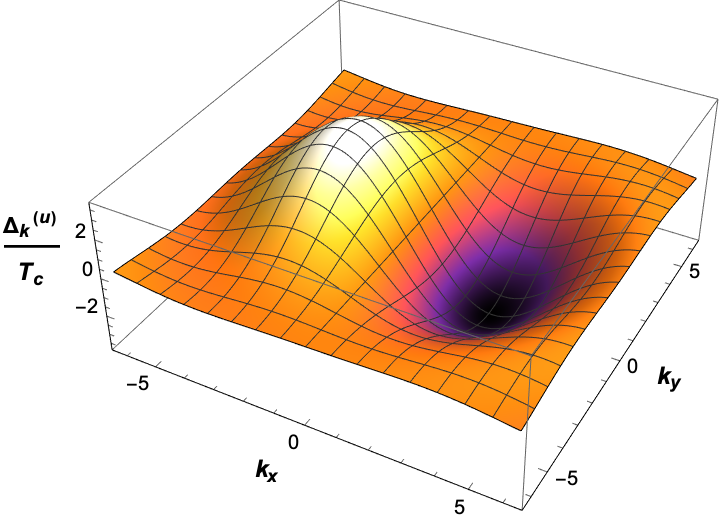}
		\caption{For $\delta=\delta_{+}=\frac{5}{4}$}
	\end{subfigure}
	\hfil
	\begin{subfigure}[b]{0.3\textwidth}
		\centering
		\includegraphics[scale=0.45]{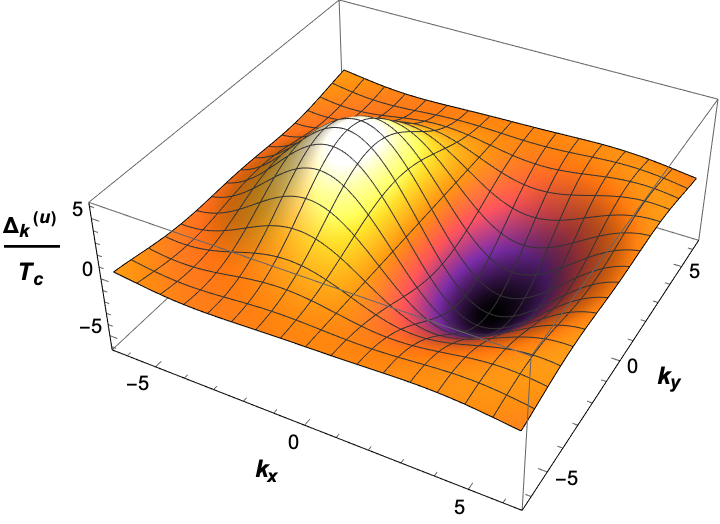}
		\caption{For $\delta=\delta_{+}=\frac{7}{4}$}
	\end{subfigure}
	\caption{The $p$-wave gap function (order parameter) for $m^2=-\frac{3}{16}$ with $T=0.9T_c$}
	\label{fignw11}
\end{figure}

We would like to mention the general prescription for
	the mapping between the order parameter (gap function) and the matter field  by the near boundary behavior of the bulk field $\psi$: 
		\begin{eqnarray}
		\psi(z, u, \theta) \sim  \langle\mathcal{O}(u,\theta ) \rangle z^{\delta-1 
		}, \quad
{\tilde\psi}(z, k, \theta) \sim \Delta_{\vec{k}}(k,\theta) z^{\delta -1} , 
	\end{eqnarray}
	in the limit $z\to 0$. 
	Here  ${\tilde\psi}(z, k, \theta)= \int d^{2}x e^{i\vec{k}\cdot \vec{x}} \psi(z, u, \theta)$ is the Fourier transform and $k=\sqrt{k^2_x+k^2_y}$.
	Since the radial coordinate in gravity theory is associated with the energy in boundary theory,    the temperature dependence is solely coming from the radial part of the field $\Psi(z)$ and angle dependence  is from the solution 
	$\mathcal{R}(u, \theta)$ so that we can see that these dependencies 
	are  factorized as in the eq. (\ref{vortexfree3}). 
	From this, the angle dependent condensation operator can be written as
	\begin{eqnarray}
		\langle\mathcal{O}\rangle=
		\Delta(T)  \mathcal{H}(u, \theta) , \quad 
		\Delta_{k}= \Delta(T)\mathcal{I}(k_x, k_y)  ,
		\label{ord2}
	\end{eqnarray} 
	where $\mathcal{H}(u, \theta)$ is $\mathcal{R}(u, \theta)$ upto a constant and    $\mathcal{I}(k_x,k_y)$ is the  Fourier transformation of $\mathcal{H}(u, \theta)$. 
 From eq.(\ref{eq.4.49}), 
we can   write the solution {\it near boundary} for ground state as
\begin{eqnarray}
	\psi_u=\tilde{\beta} T_c \sqrt{1-\frac{T}{T_c}} \cos\theta z^{\delta-1}
	 ~~&\text{and}&~~ \psi_{\theta}= -\tilde{\beta} T_c \sqrt{1-\frac{T}{T_c}} u\sin\theta z^{\delta-1},
\end{eqnarray}
from which 
\be 
\mathcal{H}_{u}(u,\theta)=\cos\theta,\quad \quad  \mathcal{H}_{\theta}(u,\theta)=-u\sin\theta,
\ee 
\begin{eqnarray}
	\Delta(T)
	= \tilde{\beta}  T_c \sqrt{1-\frac{T}{T_c}} = \left\{ 
	\begin{array}{ c l }
		7.359  ~T_c \sqrt{1-\frac{T}{T_c}} & \quad \textrm{for } \delta=\delta_{-} = \frac{5}{4}  \\
		11.96 ~ T_c \sqrt{1-\frac{T}{T_c}} & \quad \textrm{for } \delta=\delta_{+} = \frac{7}{4}
	\end{array}
	\right. ~.
\end{eqnarray}
 The  Fourier transformation of $\mathcal{H}_{u}(u,\theta) ~\text{and}~ \mathcal{H}_{\theta}(u,\theta)$ 
 gives \footnote{ Two dimensional Fourier transformation is given by
	$\mathcal{I}(k_x, k_y)= \int_{0}^{a} \int_{0}^{2\pi} \mathcal{H}(u, \theta) e^{-iu(k_x\cos\theta+ k_y\sin\theta)} u du d\theta $. }
\begin{eqnarray}
	\Delta^{(u)}_{k}= \Delta(T) \mathcal{I}_u(k_x,k_y) ~~~\text{and}~~~ \Delta^{(\theta)}_{k}= \Delta(T) \mathcal{I}_{\theta}(k_x, k_y)
\end{eqnarray}
where 
\be 
 \mathcal{I}_u(k_x,k_y) = -2\pi a^2 \frac{i ak}{6} \, _1F_2\left(\frac{3}{2};2,\frac{5}{2};\frac{-a^2 k^2}{4} \right)\cos\theta, \quad ~ \mathcal{I}_{\theta}(k_x, k_y)=2\pi a^2\frac{i J_2(a k)}{k}\sin\theta, 
 \ee
with 
 $\, _1F_2\left(\frac{3}{2};2,\frac{5}{2};\frac{-a^2 k^2}{4} \right)$ being a   hypergeometric function. Inspite of their major difference,  the gap structure from this two components are connected by the just phase factor    $\psi_{u}(r, u,\theta)$ and $\psi_{\theta}(r, u,\theta)$ are related by eq.(\ref{relutheta}). See 
 the density plots in the figure  (\ref{fig15pv}) where we draw the $p$-wave gap function for fixed value of $a$. We   now focus  on the gap function $\Delta^{(u)}_{k}$  since $\psi_{u}$ represents the order parameter of the system. The ratio   $\frac{\Delta^{(u)}_{k}}{T_c}$   is shown in figure (\ref{fignw11}) for $m^2=-\frac{3}{16}$ at $T=0.9T_c$. 

   
\subsection{Angular dependent waves     in scalar field model} 
Our task is now to ask whether the $p$-wave gap function can be  simply obtained from a scalar order model and if not, to ask  what is the 
  differences between the p-wave states in vector model and the scalar model? We examine the angle dependent scalar field in the Abelian-Higgs Model in Appendix \ref{appen1}. In this  section, we   describe    just physics of  angle dependent  wave states in the scalar field model. 

\begin{figure}[h!]
	\centering
	\begin{subfigure}[b]{0.47\textwidth}
		\centering
		\includegraphics[scale=0.4]{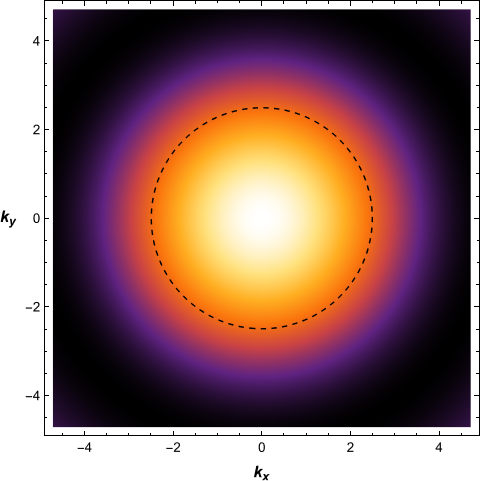}
		\caption{s-wave state ($l=0$)}
		\label{a1}
	\end{subfigure}
	\hfill
	\begin{subfigure}[b]{0.47\textwidth}
		\centering
		\includegraphics[scale=0.4]{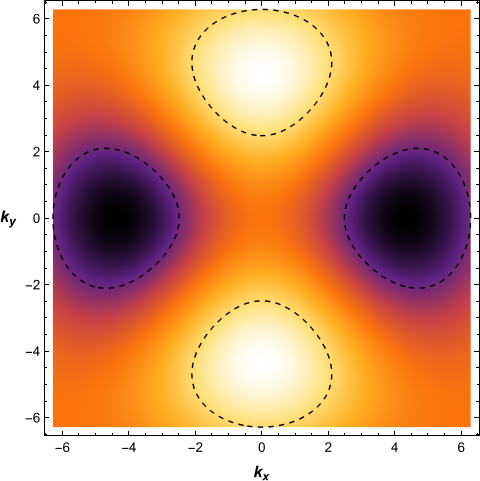}
		\caption{d-wave state ($l=2$)}
		\label{a3}
	\end{subfigure}
	\caption{Density plot of $l$-wave states in momentum space.   $\alpha_{01}=0,  \alpha_{21}=5.1356$ for  $l=0, 2$ respectively.}
	\label{fig11}
\end{figure}

When we consider the angle dependent fields in the scalar field model, two dimensional Laplacian appears in the matter field equations. Here, we are interested on the solution of the Laplacian part only in the matter field equation. After expanding   both fields, we will be able to use the separation variables method for solving the Laplacian part of the matter field equations. 
Substituting $\mathcal{R}(u, \theta)= R(u) \Theta(\theta)$ in the Laplacian part of the matter field eq.(\ref{kp019}), we obtain 
\begin{eqnarray}
	\frac{ \partial_{u}[u \partial_{u}R(u)]}{u R(u)} + \frac{1}{u^2} \frac{\partial^2_{\theta}\Theta(\theta)}{\Theta(\theta)} = -\alpha^2 ~.
	\label{kp20}
\end{eqnarray}
The angle dependence part is separated by the separation constant $l$ in which equation takes form
\begin{eqnarray}
	\frac{d^2\Theta(\theta)}{d\theta^2}+l^2 \Theta(\theta)=0 ~.
\end{eqnarray}
The solution   can be  chosen such   that  $\Theta(\theta)=  e^{il \theta}$, where $l$ can be identified as angular quantum number.  Using the separation constant $l$, the equation (\ref{kp20}) now becomes 
\begin{eqnarray}
	\frac{d^2R(u)}{du^2} +\frac{1}{u} \frac{dR}{du} +\left(\alpha^2-\frac{l^2}{u^2}\right) R(u)=0, 
\end{eqnarray}
which  is nothing but the Bessel equation whose solution takes form  
\begin{eqnarray}
	R(u)= J_{l}\left(\alpha u\right)
\end{eqnarray} 
for the finiteness at the origin $u=0$. 
The value of $u$ runs from $0$ to system size $a$. The $R(u)$ should vanish at the boundary of the system for which we have to set $\alpha=\frac{\alpha_{l1}}{a}$ where $\alpha_{l1}$ is the first zero of the $J_{l}$ polynomial. The solution now reads  
\begin{eqnarray}
	R(u)= J_{l}\left(\alpha_{l1}\frac{u}{a}\right)~.
\end{eqnarray}
We can finally express the solution $\mathcal{R}(u, \theta)$  in the following form  
\begin{eqnarray}
	\mathcal{R}(u, \theta)= R(u)\Theta(\theta)=   J_{l}\left(\frac{\alpha_{l1}}{a}u\right)e^{il \theta} . 
	\label{eq36}
\end{eqnarray} 
For the angular momentum $l=0$, the ground state is independent of $\theta$ which implies that the ground state in the scalar field model is represented by $s$-wave state. The wave state for non-zero $l$ represents the excited states in the scalar field model. 
 Since $\alpha=0$ in eq.(\ref{rd19}) recovers the field eq.(\ref{hscfe0}), we set $\alpha_{01}=0$ for $l=0$ which gives trivial solution of $\mathcal{R}(u, \theta)=1$ for $s$-wave state.
To visualize the wave     in momentum space, we now make the Fourier transformation of this solution $\mathcal{R}(u, \theta)$ which yields as follow 
\begin{eqnarray}
	\mathcal{I}(k, \phi) &=&  \int_{0}^{a} \int_{0}^{2\pi} J_{l}\left(\frac{\alpha_{l1}}{a}u\right)e^{il \theta} e^{-i ku\cos(\theta-\phi)} u du d\theta
\end{eqnarray}
where $k=\sqrt{k^2_x+k^2_y}, k_x=k\cos\phi, k_y=k\sin\phi$ and $\phi$ is the angle in momentum space. After some calculation, we obtain 
\begin{eqnarray}
	\label{delta1}
	\mathcal{I}(k, \phi)&=& 2\pi a^2    (-i)^l  e^{il\phi} \frac{ \left[\alpha_{l1} J_{l-1}(\alpha_{l1}) J_{l}(k a)- k a J_{l-1}(ka) J_{l} (\alpha_{l1}) \right]}{k^2 a^2- \alpha_{l1}^2}   
\end{eqnarray}
This result is very crucial for understanding the different wave state structures in the scalar field model.  
We can identify the      angles in coordinate space and that in momentum space,
 $\phi=\theta$  
as it is well known,  the   states in momentum space become 
\begin{eqnarray}
	\mathcal{I}(k, \theta) = \left\{ 
	\begin{array}{ c l }
		2\pi a^2 ~ \frac{J_1(ka)}{ka} & \quad \textrm{for } l = 0 
		\\
		2\pi a^2 ~(-i)^l e^{i l\theta}~ \frac{ \alpha_{l1} J_{l-1}(\alpha_{l1}) J_l(k a)}{k^2 a^2- \alpha_{l1}^2}                & \quad \textrm{otherwise}
	\end{array}
	\right. ~~.
	\label{eq41}
\end{eqnarray} 
The $\alpha_{l1}$  values are  $\alpha_{01}=0, \alpha_{11}=3.8317, \alpha_{21}=5.1356$ for $s, p, d$-wave state respectively. Using real part of $\mathcal{I}(k, \theta)$ (\ref{eq41}), the density plot of $s, d$-wave states in momentum space are presented in Figure \ref{fig11}, where $k_x$ and $k_y$ is expressed in inverse unit of $a$. 

\section{Comparing the scalar field vs the vector field  models} 
\subsection{The critical temperatures  }
\noindent{ We already know that the ground state in scalar   model is in $s$-wave state and the ground state in vector field model is in $p$-wave state. In this subsection, we will discuss   the critical temperature of the ground state   in both models and compare   them for each value of scaling dimension $\delta$. For example, at   the same value of the scaling dimension $\delta=2$, the critical temperature for vector field is $T_c=0.124\sqrt{\rho}$  while that of  the scalar field model $(T_c=0.117\sqrt{\rho})$ so that 
\be
T_{c}^{p-wave}>T_{c}^{s-wave}, 
\ee
This   interesting result continue to hold for other values of $\delta$. 
The critical temperature for $p$-wave state matches with the results of non-abelian holographic model.  See figure (\ref{fignw10}).
\begin{figure}[h!]
	\centering
		\includegraphics[scale=0.8]{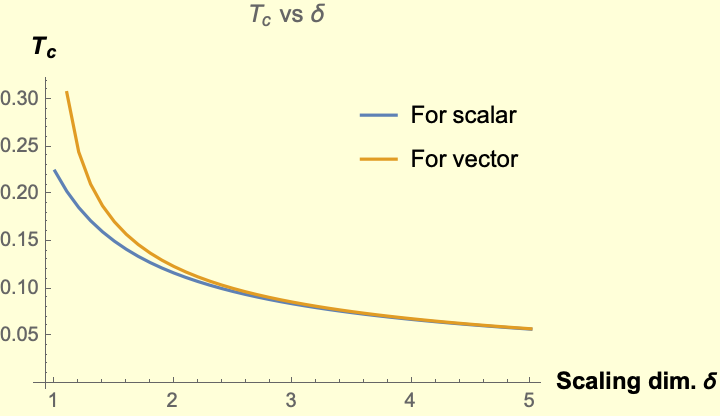}
	\caption{Comparing $T_{c}$ in the Vector vs Scalar models}
\label{fignw10}
\end{figure} 
The difference in the critical temperature of ground state in both models is mainly because of the difference in $q(z)$ in Sturm-Liouville form whereas $p(z), r(z)$ are same in both models. For the ground state in scalar field model, we can write $q_s(z)=z^{2\delta-2}g(z)\left[\frac{\delta(\delta-3)}{z^2}+\frac{g'(z)\delta}{zg(z)}-\frac{m^2}{z^2 g(z)}\right]$ if we recast eq.(\ref{afy5}) in  Sturm-Liouville form.
If we denote $q(z)$ as $q_v(z)$ (from eq.(\ref{i1})) for vector field model, then the difference 
\be q_v(z)-q_s(z)=z^{2\delta-1}, \label{deltaq}\ee
 which leads lower $\tilde{\lambda}$ values in vector field model. This is the mathematical reason for higher $T_c$ in vector field model. The possible physical reason for this interesting feature in holographic setup may be lurk in the instability of the bulk field since mass of the fields are different for same value of the scaling dimension. The mass of the scalar field and the vector field are $m^2_s=\delta(\delta-3)$ and $m^2_v=(\delta-1)(\delta-2)$ respectively, they are related by 
\be
m^2_v=m^2_s+2,
\ee
which is responsible for the simple result of eq.(\ref{deltaq}).
Before we finish this subsection, we mention that in the ref. \cite{caispcom}, the competition between the s-and p-wave condensations were studied.  However, the authors compared p-wave and s-wave such that the p-wave model has fixed conformal weight 2 while the s-wave model has varying weights. In contrast, we compared s-wave and p-wave at the same weight for various values of weight. In the presence of the condensate and the charge density, it is not necessary to respect the Lorentz invariance and density operator and current operator may have different weights.

\subsection{Comparing  p-wave states in vector and scalar  models}
The $p$-wave state in the scalar field model is in the excited state of the system  while that   state in the vector field model  is the ground state. Nevertheless they can be the same since they are states in different models. Therefore the question here is how much they are different if they are different.  {Before we compare these, we would like to mention that the condensation to p-wave state in the scalar field model is possible only under  the constraint such that $s$-wave condensation is forbidden for some reason. In such situation, the $p$-wave state is the ground state in the scalar field model. Then one may ask whether the $p$-wave gap structure in scalar field model and in vector field model are similar or not.}
We only need to focus on the momentum dependent part $\mathcal{I}(k, \theta)$ of the gap function here. From the scalar field model, the excited $p$-wave state in momentum space is represented by (from eq.(\ref{eq41}))
\begin{eqnarray}
	\mathcal{I} (k, \theta)=- 2\pi a^2 i \frac{\alpha_{11}J_{0}(\alpha_{11}) J_1(k a)}{ k^2 a^2-\alpha^2_{11}} e^{i \theta } 
	\label{com1}
\end{eqnarray}
\begin{figure}[h!]
	\centering
	\begin{subfigure}[b]{0.3\textwidth}
		\centering
		\includegraphics[scale=0.45]{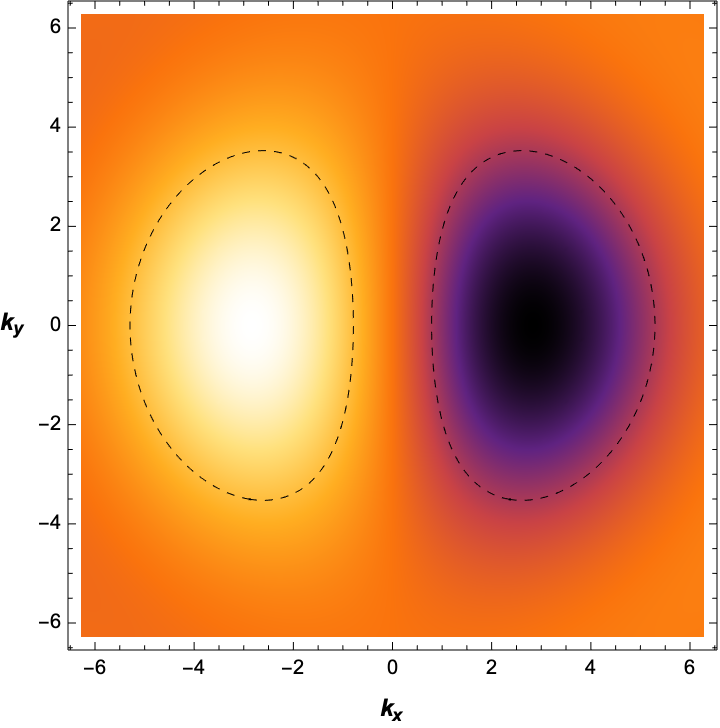}
		\caption{Imaginary part of $\mathcal{I} (k, \theta)$}
		\label{a1111pv1}
	\end{subfigure}
	\hfil
	\begin{subfigure}[b]{0.3\textwidth}
		\centering
		\includegraphics[scale=0.45]{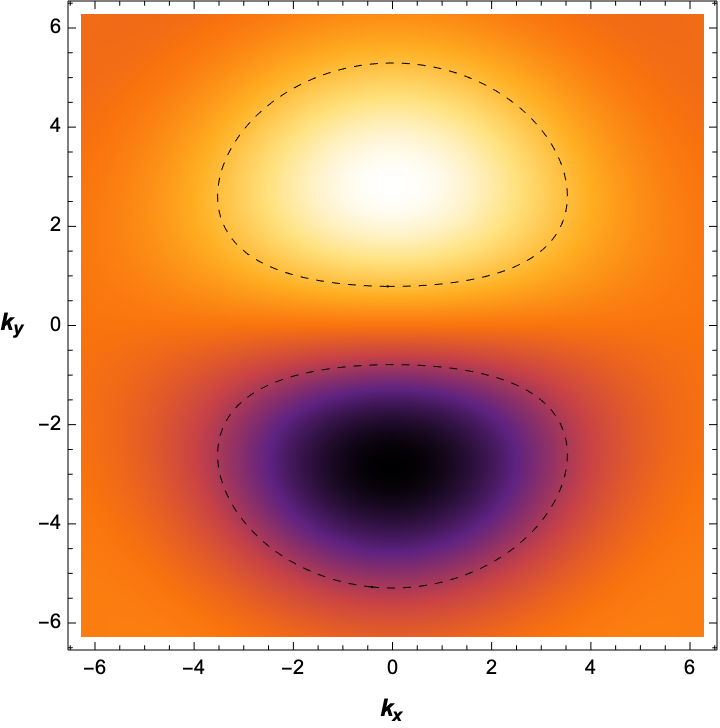}
		\caption{Real part of $\mathcal{I} (k, \theta)$}
		\label{a2111pv1}
	\end{subfigure}
	\caption{The excited $p$-wave state from the scalar field model.}
	\label{pfig}
\end{figure} 
The momentum dependence part of the gap energy in vector field model reads
\begin{eqnarray}
\mathcal{I}_u(k, \theta) = -2\pi a^2 \frac{i ak}{6} \, _1F_2\left(\frac{3}{2};2,\frac{5}{2};\frac{-a^2 k^2}{4} \right)\cos\theta~~\text{and}~~ \mathcal{I}_{\theta}(k, \theta)=2\pi a^2\frac{i J_2(a k)}{k}\sin\theta 
\label{com2}
\end{eqnarray}
Since $\mathcal{I}_u(k, \theta) $ and $\mathcal{I}_{\theta}(k, \theta)$ are related and $\psi_u$ is the measure of the order parameter, we now discuss only about the $\mathcal{I}_u(k, \theta) $ part from the gap energy $\Delta_{k}^{(u)}$.  
From the Fig.(\ref{fig15pv}) and Fig.(\ref{pfig}), we observe that the density plot of the imaginary part of $\mathcal{I}(k,\theta)$ is very similar to the density plot of $\mathcal{I}_u(k,\theta)$. 
We now take the ratio between this two function 
\begin{eqnarray}
	\frac{Im.(\mathcal{I}(k,\theta))}{\mathcal{I}_u(k,\theta)} = \frac{\alpha_{11}J_{0}(\alpha_{11}) J_1(k a)}{ (k^2 a^2-\alpha^2_{11})ka} \frac{6}{ \, _1F_2\left(\frac{3}{2};2,\frac{5}{2};\frac{-a^2 k^2}{4} \right)}
	\label{com3}
\end{eqnarray}
which is independent of $\theta$. We can now plot this ratio function (\ref{com3}) in Figure \ref{comfig} for fixed value of $a$, which is almost constant function for $k a<<1$ \footnote{If we consider $k a<1$, then we can write this ratio function as
		$\frac{Im.(\mathcal{I}(k,\theta))}{\mathcal{I}_u(k,\theta)} \approx 0.32 + 0.006 k^2 a^2$.
In this limit $k a<1$, we can neglect the higher order terms in $k^2a^2$.}. Notice , however, that this ratio function diverges at the   roots of the generalized hypergeometric function. The first of which is $k a=5.8843$.  
\begin{figure}[h!]
	\centering
			\begin{subfigure}[b]{0.25\textwidth}
			\centering
			\includegraphics[scale=0.45]{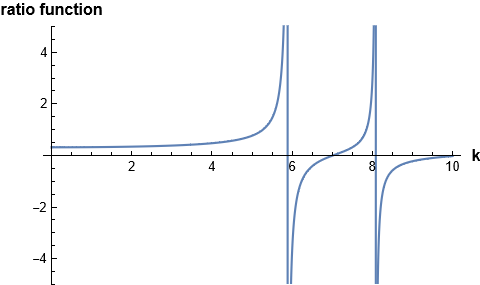}
			\caption{}
			\label{2dfig}
		\end{subfigure}
		\hfil
		\begin{subfigure}[b]{0.25\textwidth}
			\centering
			\includegraphics[scale=0.45]{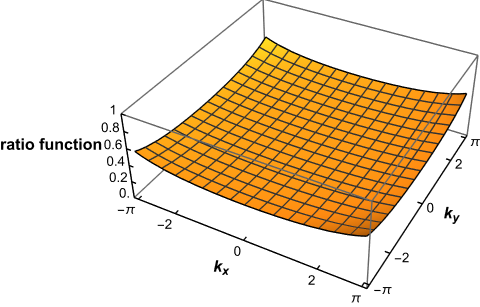}
			\caption{}
			\label{3dfig}
		\end{subfigure}
	\caption{The plot of the ratio function}
	\label{comfig}
\end{figure} 

\section{Discussion}
In this paper, we have investigated the angle dependent gap structures in the vector field models. In order to understand the necessity of  the vector field model for $p$-wave superconductors, we have started with the vector field model.  We  showed that the normalizable ground states of this system is  given by the   $p$-wave state while  the state with $l_p=0$ is not normalizable.  Therefore,   the $p$-wave ground state can be achieved only from the vector field model.  We have found that the order parameter for vector field model is represented by $\psi_{u}$, not the $\psi_x$ since $\psi_x$ does not have any angular dependency. 

We then explore the angular dependence  in the scalar field model, 
where all  $l$-wave states are available. Here the ground state of the system is from the $s$-wave state. For $l >0$, they are excited states of the system.    We have then compare the momentum dependent part $\mathcal{I}_{k,\theta}$ of gap function from the scalar field model and the vector field model. We observe that the structure of both gap energy is almost same  for small momentum range.
 The point is that $p$- and $d$-wave gap structures can be explained through the     scalar field model   if we  assume that the states for lower value of angular momentum number $(l=0,1)$ is forbidden in the scalar field model}.

We also studied 
the critical temperature   in the probe approximation of the gravity background using matrix-eigenvalue algorithm method and Sturm-Liouville's eigenvalue method for different scaling dimensions.  Another interesting point is that the critical temperature for the ground state of vector field model is higher than the ground state of the scalar field model for same value of the scaling dimensions. 

 We would now like to mention the drawbacks and future works.
 The Fermi surface is not easily demonstrated  in our   set-up, where  fermions are not included at all.  
The appearance of Fermi arc and Fermi surface is only possible when one consider the interaction between fermion and tensor field in holographic set-up\cite{dhsc12}. 
We will come back to this issue in the future work.

\section*{Acknowledgments} DG would like to thank Taewon Yuk for various discussions and for helping in using the Mathematica. 
This  work is supported by Mid-career Researcher Program through the National Research Foundation of Korea grant No. NRF-2021R1A2B5B02002603. 
We  thank the APCTP for the hospitality during the focus program, where part of this work was discussed.

\appendix
\section{Abelian-Higgs model with angular dependent scalar field}
\label{appen1}
\begin{figure}
	\centering
	\begin{subfigure}[b]{0.45\textwidth}
		\centering
		\includegraphics[scale=0.5]{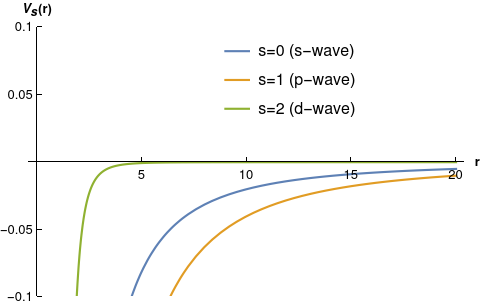}
		\caption{$V_s(r)$ from generalized spin field model}
		\label{a200}
	\end{subfigure}
	\hfill
	\begin{subfigure}[b]{0.45\textwidth}
		\centering
		\includegraphics[scale=0.5]{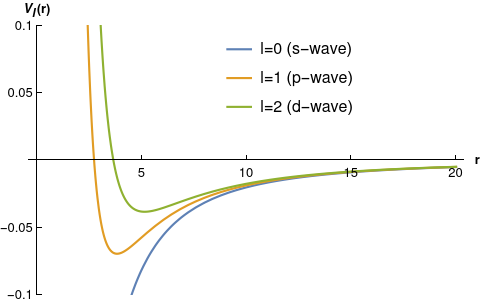}
		\caption{$V_l(r)$ from scalar field model}
		\label{a300}
	\end{subfigure}
	\caption{Plot of $V_s(r)$ and $V_l(r)$ for $m^2=-2$ with Schwarzschild-AdS$_4$ metric with $r_{h}=1$ in different wave states $s, p, d$. Left figure is for spin field model with different spin value and the figure in right hand side is for Abelian-Higgs model with $\alpha=\frac{\alpha_{l1}}{a}=0, 3.8317, 5.1356$ (setting $a=1$) for different angular momentum value $l=0, 1, 2$ respectively. Here $l=1$ and $l=2$ represent the excited states in the scalar field model.}
	\label{fig10}
\end{figure}

In the scalar order model, angular dependent matter field and gauge ansatz are
\begin{eqnarray}
	\psi=\psi(r, u, \theta) ~~,~~~~~ 	A = A_t(r, u, \theta)dt ~~
	\label{vector0}
\end{eqnarray}
Using the fields ansatz (\ref{vector0}) in fields eq.(\ref{ac1}), we obtain the gauge field and the scalar field equation
\begin{scriptsize}
	\begin{eqnarray}
		\label{phio}
		\partial_{r}[r^2\partial_{r}A_t(r, u, \theta)]+ \frac{1}{f(r)} \left[ \frac{\partial_{u}[u \partial_{u}A_t(r, u, \theta)]}{u} + \frac{\partial^2_{\theta}A_t(r, u, \theta)}{u^2}\right]&=&  \frac{2q^2r^2}{f(r)} |\psi(r, u, \theta)|^2 A_t(r, u, \theta) \\
		\partial_{r}[r^2 f(r)\partial_{r}\psi(r, u, \theta)]+ \left[\frac{\partial_{u}[u \partial_{u}\psi(r, u, \theta)]}{u}+ \frac{\partial^2_{\theta}\psi(r, u, \theta)}{u^2}\right] &=& r^2\left[-\frac{q^2A^2_t(r, u, \theta)}{f(r)}+ m^2\right]\psi(r, u, \theta) ~. 
		\label{psio}
	\end{eqnarray}
\end{scriptsize}
\noindent Because of the non-linear coupling term, we can not use the separation variables technique to solve eq.(s)(\ref{phio},\ref{psio}). We therefore expand the both field as a series in a small parameter $\epsilon$:
\begin{eqnarray}
	A_t(r, u, \theta) &=& \mathcal{A}^{(0)}_t(r)  + \epsilon A^{(1)}_t(r, u, \theta)+ ... \\
	\psi(r, u, \theta) &=& \Psi^{(0)}(r) + \sqrt{\epsilon} \psi^{(1)}(r, u, \theta) + ...  ~~.
\end{eqnarray}
Comparing the power of $\epsilon$ in fields equations (\ref{phio}, \ref{psio}), we obtain for gauge field
\begin{eqnarray}
	\label{phi0}
	\epsilon^0: && \frac{d^2\mathcal{A}^{(0)}_t(r)}{dr^2} +\frac{2}{r} \frac{d\mathcal{A}^{(0)}_t(r)}{dr}= \frac{2q^2}{f(r)} |\Psi^{(0)}(r)|^2 \mathcal{A}^{(0)}_t(r) \\
	\epsilon^1: && \partial^2_{r}A_t^{(1)}(r, u, \theta)+ \frac{2}{r}\partial_{r}A_t^{(1)}(r, u, \theta) + \frac{1}{r^2 f(r)} \left[\partial^2_{u}A_t^{(1)}(r, u, \theta)+ \frac{\partial_{u}A_t^{(1)}(r, u, \theta)}{u} \right.  \nonumber \\
	&&~~~~~\left. + \frac{\partial^2_{\theta}A_t^{(1)}(r, u, \theta)}{u^2}\right] = \frac{2q^2}{f(r)} \left\{ |\psi^{(1)}(r, u, \theta)|^2 \mathcal{A}^{(0)}_t(r)+|\Psi^{(0)}(r)|^2 \mathcal{A}^{(1)}_t(r, u, \theta) \right\} ~~~~~~~
	\label{phi1}
\end{eqnarray}
and  those for the matter field
\begin{eqnarray}
	\label{psi0}
	\epsilon^{0}: && \frac{d^2\Psi^{(0)}(r)}{dr^2} +\left[\frac{\partial_r f(r)}{f(r)}+\frac{2}{r}\right] \frac{d\Psi^{(0)}(r)}{dr}  =\left[-\frac{q^2\mathcal{A}^{(0)2}_t(r)}{f^2(r)}+ \frac{m^2}{f(r)}\right]\Psi^{(0)}(r) \\
	\sqrt{\epsilon}: && \partial^2_{r}\psi^{(1)}(r, u, \theta)+ \left[\frac{\partial_r f(r)}{f(r)}+\frac{2}{r}\right]\partial_{r}\psi^{(1)}(r, u, \theta) + \frac{1}{r^2 f(r)}\left[\partial^2_{u}\psi^{(1)}(r, u, \theta)\right. \nonumber \\
	&&~~\left. + \frac{ \partial_{u}\psi^{(1)}(r, u, \theta)}{u}+ \frac{\partial^2_{\theta}\psi^{(1)}(r, u, \theta)}{u^2}\right]= \left[-\frac{q^2\mathcal{A}^{(0)2}_t(r) }{f^2(r)}+ \frac{m^2}{f(r)}\right]\psi^{(1)}(r, u, \theta) ~.~~~
	\label{psi1}
\end{eqnarray}
Since we expand the fields at near $T_c$, we can identify the small parameter 
$\epsilon=1-\frac{T}{T_c}$.
Since $\psi(r, u, \theta)=0$ at $T=T_c$ and at the nodes, we have to set $\Psi^{(0)}(r)=0$. Therefore, the matter field can be expressed as 
\begin{eqnarray}
	\psi(r, u, \theta)=\psi^{(1)}(r, u, \theta)\sqrt{1-\frac{T}{T_c}}~.
	\label{ord1}
\end{eqnarray}
We now need to solve $\psi^{(1)}(r, u, \theta)$ in order to know the gap structure since $\psi(\infty, u, \theta)$ is the order parameter of the boundary theory. 
Using the separation of variables method in eq.(\ref{psi1}), scalar field can be separate out as 
\begin{eqnarray}
	\psi^{(1)}(r, u, \theta) =\Psi^{(1)}(r)\mathcal{R}(u,\theta) ~.
\end{eqnarray} 
With this, the scalar field equation now becomes
\begin{eqnarray}
	\frac{1}{\Psi^{(1)}(r)} \left[\frac{d^2\Psi^{(1)}(r)}{dr^2} + \left\{\frac{\partial_r f(r)}{f(r)}+\frac{2}{r}\right\} \frac{d\Psi^{(1)}(r)}{dr} \right] + \left\{\frac{q^2A^{(0)2}_t(r)}{f^2(r)}- \frac{m^2}{f(r)}\right\} \nonumber \\
	=- \frac{1}{r^2f(r)} \frac{1}{\mathcal{R}(u, \theta)} \left[\partial^2_{u}\mathcal{R}(u, \theta) +\frac{\partial_{u}\mathcal{R}(u, \theta)}{u}+ \frac{\partial^2_{\theta}\mathcal{R}(u, \theta)}{u^2} \right] :=\alpha^{2}~.
	\label{kp019}
\end{eqnarray}
Right hand side of the above equation is two dimensional Laplacian in $u, \theta$ coordinate. 
Using the separation constant $\alpha^2$, the radial part of the matter field becomes
\begin{eqnarray}
	\frac{d^2\Psi^{(1)}(r)}{dr^2} + \left\{\frac{\partial_r f(r)}{f(r)}+\frac{2}{r}\right\} \frac{d\Psi^{(1)}(r)}{dr}  + \left\{\frac{q^2A^{(0)2}_t(r)}{f^2(r)}- \frac{m^2}{f(r)} -\frac{\alpha^2}{r^2f(r)}\right\}\Psi^{(1)}(r) &=& 0 ~.~~~~~
	\label{rd19}
\end{eqnarray}
For $\alpha=0$, the above equation becomes same as eq.(\ref{hscfe0}) which tells us that the trivial solution of $\mathcal{R}(u,\theta)$=constant represents the solution of field without any angular dependency. The non-zero value of $\alpha$ gives us the excited states for the scalar field model.  
The value of separation constant $\alpha=\frac{\alpha_{l1}}{a}$ is determined by the first root the Bessel functions with angular quantum number $l$, where $a$ is the system size (see section 5).
We can now recast the radial part of the scalar field in the following form
\begin{eqnarray}
	\frac{d^2\Psi^{(1)}(r)}{dr^2} + \left\{\frac{\partial_r f(r)}{f(r)}+\frac{2}{r}\right\} \frac{d\Psi^{(1)}(r)}{dr}  + \left\{\frac{q^2A^{(0)2}_t(r)}{f^2(r)}- V_l(r) \right\}\Psi^{(1)}(r) = 0 ~~~
	\label{mfe2}
\end{eqnarray}
where the effective potential, $V_l(r)$,  is given by 
\begin{eqnarray}
	V_l(r) = \frac{m^2}{f(r)} +\frac{\alpha^2}{r^2f(r)}~~.
	\label{vlf}
\end{eqnarray}
We would like to mention that some feature of $V_l(r)$    are quite different from the  ``effective potential" of the ground states in (\ref{spin3}), 
which has no centrifugal potential $1/r^{2}$  for $s=0,1$. 
We will need to consider the vector field ($s=1$) model  and tensor field  model $(s=2)$ to get the ground state of the $p$-wave and $d$-wave superconductors respectively, simply because the higher values of $l$ in scalar field model represent the excited states of the system. 
In the Figure \ref{fig10},  the ``effective potential" $V_s(r)$ (\ref{spin3}) and $V_{l}(r)$ (\ref{vlf}) are shown 
for $m^2=-2$   for different $l$-wave states. The figure in right hand side in Fig.\ref{fig10} reveals similar physics with hydrogen like atom.

\subsection{The critical temperature }
The zeroth order and first order gauge field equations
\begin{eqnarray}
 \frac{d^2\mathcal{A}^{(0)}_t(r)}{dr^2} &+&\frac{2}{r} \frac{d\mathcal{A}^{(0)}_t(r)}{dr} = 0 \\
 \partial^2_{r}A_t^{(1)}(r, u, \theta)+ \frac{2}{r}\partial_{r}A_t^{(1)}(r, u, \theta) + \frac{\nabla^2_{(2)}A_t^{(1)}(r, u, \theta)}{r^2 f(r)} &=& \frac{2q^2}{f(r)} |\Psi^{(1)}(r)\mathcal{R}(u, \theta)|^2 \mathcal{A}^{(0)}_t(r)  ~~~~~~~
	\label{phi1}
\end{eqnarray}
where $\nabla^2_{(2)}$ is the two dimensional Laplacian.
The asympotic solution of the zeroth order gauge field and the radial part of the first order matter field read \cite{hs6a}
\begin{eqnarray}
\mathcal{A}^{(0)}_t=\mu - \frac{\rho}{r}	 ~~~, ~~~ \Psi^{(1)}(r) = \frac{\Psi_{-}}{r^{\delta_{-}}} +\frac{\Psi_{+}}{r^{\delta_{+}}}
\end{eqnarray}
where $\delta_{\pm}= \frac{1}{2}[3\pm \sqrt{9+4m^2}]$ is the scaling dimension in the scalar field model which is different from the vector field model.
In $z=\frac{r_h}{r}$ coordinate, zeroth order gauge field and the radial part of the first order matter field yield with the identification of $\alpha=\frac{\alpha_{l1}}{a}$
\begin{small}
	\begin{eqnarray}
		\label{gfe5}
		\frac{d^2\mathcal{A}^{(0)}_t(z)}{dz^2} &=& 0~~\\
				\frac{d^2\Psi^{(1)}(z)}{dz^2} + \left(\frac{g^{\prime}(z)}{g(z)}-\frac{2}{z}\right) \frac{d\Psi^{(1)}(z)}{dz} + \left[\frac{q^2  (\mathcal{A}^{(0)}_t(z))^2}{r^2_{h} g^2(z)}- \frac{m^2}{z^2g(z)}-\frac{\alpha^2_{l1} }{a^2 r^2_h g(z)} \right]\Psi^{(1)}(z)&=&0 ~~~~~
		\label{gfenw2}
	\end{eqnarray}
\end{small}
The first order gauge field becomes 
\begin{eqnarray}
\partial^2_{z}A_t^{(1)}(z, u, \theta) + \frac{1}{r^2_h g(z)} \nabla^2_{(2)}A_t^{(1)} (z, u, \theta) = \frac{2q^2}{z^2g(z)}  \left(\Psi^{(1)}(z)\right)^2 J_l\left(\frac{\alpha_{l1}}{a}u\right)^2 \mathcal{A}^{(0)}_t(z). 
	\label{gfenw1}
\end{eqnarray}
To estimate the critical temperature, we just need to solve the zeroth order gauge field equation (\ref{gfe5}) and first order scalar field equation (\ref{gfenw2}). At the critical temperature $T=T_c$,
the solution of the zeroth order gauge field yields
\begin{eqnarray}
	\mathcal{A}^{(0)}_t(z) = \lambda r_{h(c)} \left(1- z\right) 
	\label{gfs1}
\end{eqnarray} 
where $\lambda$ will be computed from the matter field equation using Matrix-eigenvalue algorithm.
We now substitute $\mathcal{A}^{(0)}_t(z)$ (\ref{gfs1}) in the first order scalar field equation (\ref{gfenw2}), we obtain
\begin{small}
	\begin{eqnarray}
		\frac{d^2\Psi^{(1)}(z)}{dz^2} + \left(\frac{g^{\prime}(z)}{g(z)}-\frac{2}{z}\right) \frac{d\Psi^{(1)}(z)}{dz} - \left[\frac{m^2}{z^2g(z)}+\frac{\alpha^2_{l1} \lambda}{a^2 \rho g(z)} \right]\Psi^{(1)}(z)+ \frac{q^2 \lambda^2(1-z)^2}{g^2(z)}\Psi^{(1)}(z)=0 ~~~~
		\label{afy5}
	\end{eqnarray}
\end{small}
\noindent Factoring out the behavior near the boundary $z=0$ and the horizon, we define
\begin{equation}
	\Psi^{(1)}(z)=\frac{\mathcal{C}_s}{r^{\delta}_h} z^{\delta} F(z) \hspace{1cm}\mbox{where}\;\;F(z)=
	(z^2+z+1)^{-\tilde{\lambda}/\sqrt{3}} y(z) 
	\label{eq:11}  
\end{equation}   
where $\tilde{\lambda} = q \lambda$. Then, $F$ is normalized as $F(0)=1$. We now substitute this in the matter field equation (\ref{afy5}) which yields 
\begin{eqnarray}
	&& \frac{d^2 y }{d z^2} +\frac{(1-\frac{4}{\sqrt{3}} \tilde{\lambda} +2\delta)z^3+\frac{2\tilde{\lambda}}{\sqrt{3}}z^2 +\frac{2\tilde{\lambda}}{\sqrt{3}}z+2(1-\delta)}{z( z^3-1)}\frac{d y}{d z}   	\label{eq:12}  \\
	&&+\frac{(3\delta^2-4\sqrt{3}\delta\tilde{\lambda} +4\tilde{\lambda}^2)z^2-(-3 h^2 \tilde{\lambda} + 4\tilde{\lambda}^2-2\sqrt{3}\delta \tilde{\lambda}+\sqrt{3}\tilde{\lambda})z-2\sqrt{3}(1-\delta)\tilde{\lambda}}{3 z(
		z^3-1)}y=0, 
	\nonumber
\end{eqnarray}
where $h =\frac{\alpha_{l1}}{a\sqrt{q \rho}}$.
Notice that this is the generalized Heun's equation  that has five regular singular points at $z=0,1,\frac{-1\pm\sqrt{3}i}{2},\infty$.
Substituting $y(z)= \sum_{n=0}^{\infty } d_n z^{n}$ into (\ref{eq:12}), we obtain the following four term  recurrence relation:
\begin{equation}
	\alpha_n\; d_{n+1}+ \beta_n \;d_n + \gamma_n \;d_{n-1}+\delta_n\;d_{n-2}=0  \quad
	\hbox{  for  } n \geq 2, \label{eq:13}
\end{equation}
with
\begin{equation}
	\begin{cases} \alpha_n=-3(n+1)(n+2\delta-2) \cr
		\beta_n=2\sqrt{3}\tilde{\lambda}(n+\delta-1) \cr
		\gamma_n=\sqrt{3}(2n+2\delta-3)\tilde{\lambda}+3 h^2 \tilde{\lambda} -4\tilde{\lambda}^2 \cr
		\delta_n=3(n-\frac{2}{\sqrt{3}}\tilde{\lambda}+\delta-2)^2
	\end{cases}
	\label{eq:14}
\end{equation}
The first four $d_{n}$'s are given by   $\alpha_0 d_1+ \beta_0 d_0=0$, $\alpha_1 d_2+ \beta_1 d_1+ \gamma_1 d_0=0 $, $d_{-1}=0$ and $d_{-2}=0$. 
The series  $y(z)= \sum_{n=0}^{\infty } d_n z^{n}$ is absolutely convergent for $|z|<1$. The condition for  convergence at $|z|=1$ involves parameters of the equation. 
The convergence of the series $y(z)= \sum_{n=0}^{\infty } d_n z^{n}$ can be analyzed by studying asymptotic behaviour of the linear difference equation eq. (\ref{eq:13}) as $n \rightarrow \infty$. One finds that eq. (\ref{eq:13}) possesses  three linearly independent asymptotic solution of the form
\begin{equation}
	\begin{cases}  d_1(n)\sim n^{-1}   \cr
		d_2(n)\sim  \left( \frac{-1+\sqrt{3}i}{2} \right)^n  n^{-1-\frac{2\tilde{\lambda}}{\sqrt{3}}}  \cr
		d_3(n)\sim  \left( \frac{-1-\sqrt{3}i}{2} \right)^n n^{-1-\frac{2\tilde{\lambda}}{\sqrt{3}}} 
	\end{cases}
	\label{eq:32}
\end{equation}
$d_2(n)$ and  $d_3(n)$  are called  minimal solutions to eq. (\ref{eq:13}),  and  $d_1(n)$ represents a dominant one \cite{Jone1980}. This distinction reflects the property 
$	\lim_{n\rightarrow\infty} { d_2(n)}/{ d_1(n)}=\lim_{n\rightarrow\infty} { d_3(n)}/{ d_1(n)}=0,$
because $\tilde{\lambda} >0$. 
Now we ask  when the series converges at the boundary point $z=1$. It has been known \cite{Ronv1995} that we have a convergent solution of $y(z)$ at $|z|=1$ if only if the four term recurrence relation Eq.(\ref{eq:13}) has a minimal solution. 
According to Pincherle’s Theorem  \cite{Jone1980},
$(d_n)_{n\in \mathbb{N}}$ is the minimal solution  if   
$
\alpha_0 \ne 0  $
and 
\begin{equation}
	det\left(M_{N\times N}\right)= \begin{vmatrix}
		\beta_0& \alpha_0&  &  &  &  &  &     \\
		\gamma_1 & \beta_1  &  \alpha_1 &  &  &  &  &    \\
		\delta_2  & \gamma_2  & \beta_2  & \alpha_2  &  &  &  &     \\
		& \delta_3 & \gamma_3   & \beta_3 &   \alpha_3 &  &  &    \\
		&  & \delta_4 &  \gamma_4 & \beta_4 &  \alpha_4 &  &     \\
		&  &  & \ddots & \ddots & \ddots & \ddots &     \\
		&  &  &  &\delta_{N-1}  &\gamma_{N-1}  &\beta_{N-1}  &\alpha_{N-1}   \\
		&  &  &  &  & \delta_{N} & \gamma_{N} & \beta_{N}
		
	\end{vmatrix} =0, 
	\label{eq:24}
\end{equation}
in the limit   $N\rightarrow \infty$. 
One should remember that $\alpha_{n},\beta_{n},\gamma_{n},\delta_{n}$'s are functions of $\tilde{\lambda} $
so that eigenvalues  are the solution of the above equation.   Notice also that 
Eq.(\ref{eq:24})   becomes a polynomial of degree $N$ with respect to $\tilde{\lambda} $. 
To find $\tilde{\lambda} $ for a given $\delta$, we should increase $N$ until  roots $\tilde{\lambda} $  become constant to within the desired precision \cite{Leav1990}. 

\begin{figure}[h!]
	\centering
	\begin{subfigure}[b]{0.4\textwidth}
		\centering
		{\includegraphics[width=\textwidth]{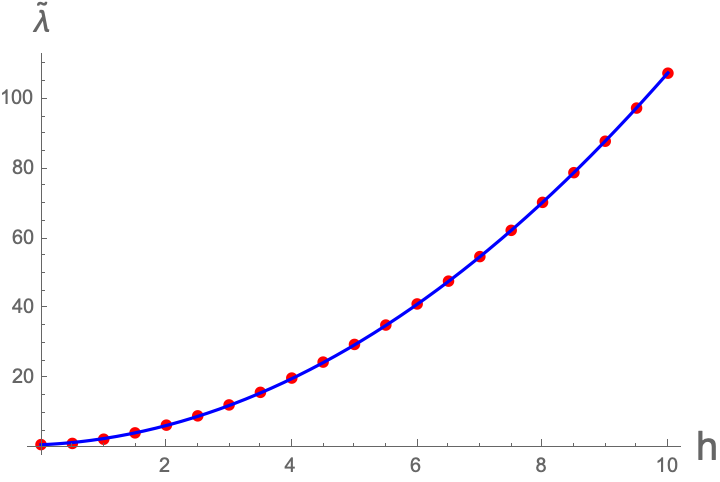}}
		\caption{For  $\delta=1$}
		\label{a11y01}
	\end{subfigure}
	\hfill
	\begin{subfigure}[b]{0.4\textwidth}
		\centering
		{\includegraphics[width=\textwidth]{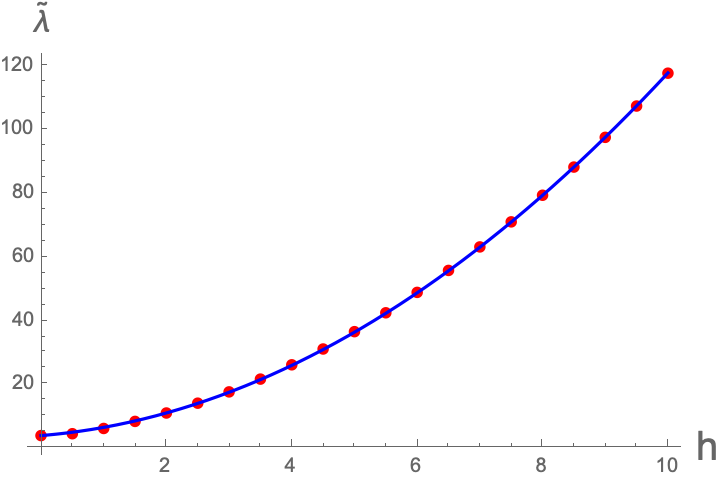}}
		\caption{For $\delta=2$}
		\label{a15y01}
	\end{subfigure}
	\caption{For numerical $\lambda$ values with the fitting function in terms of $h$ for $\delta=1, 2$ cases.}
	\label{figy}
\end{figure}

\begin{figure} 
	\centering
	\begin{subfigure}[b]{0.3\textwidth}
		\centering
		{\includegraphics[width=\textwidth]{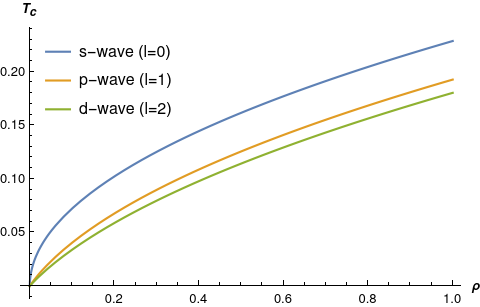}}
		\caption{For $a=10$ and For $\delta=1$}
		\label{a15y0}
	\end{subfigure}
	\hfill
	\begin{subfigure}[b]{0.3\textwidth}
		\centering
		{\includegraphics[width=\textwidth]{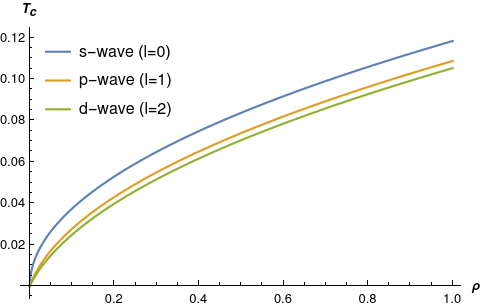}}
		\caption{For $a=10$ and For $\delta=2$}
		\label{a15y}
	\end{subfigure}
	\hfill
	\begin{subfigure}[b]{0.3\textwidth}
		\centering
		{\includegraphics[width=\textwidth]{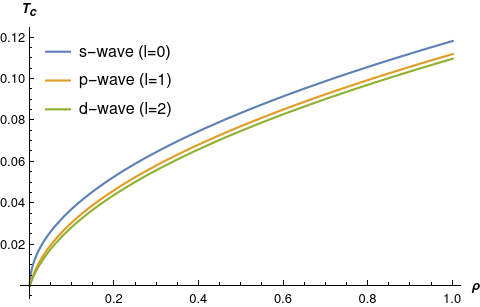}}
		\caption{For $a=15$ and For $\delta=2$}
		\label{a110y}
	\end{subfigure}
	\caption{$T_{c}$ plot in term of charge density for different sample sizes $a=(10, 15)$ for $ l=0 (s\text{-wave}), l=1 (p\text{-wave}), l=2 (d\text{-wave})$ states with different $\delta$ of the scalar field model.}
	\label{fig100}
\end{figure}

For computation of roots, we choose $N=30$. For given value of $h$, we have numerically solved the eigenvalue $\tilde{\lambda}$ using the above mentioned procedure. The smallest positive real roots of the $\tilde{\lambda}$ is corresponding to the ground state of the system. Using our numerical results for the smallest positive real roots of $\tilde{\lambda} $ and approximate fitting function, we  find the $\lambda$ values in terms of $h$, which takes in the following form
\begin{eqnarray} 
	\tilde{\lambda}  &\approx& 0.987 h^2 + 0.784 h + 1.09  \;\;\mbox{for}\;\;   \delta=1~ \\
	\tilde{\lambda}  &\approx&  0.979 h^2 + 1.591 h + 4.069 \;\;\mbox{for}\;\;   \delta=2 ~.
	\label{qq:37}
\end{eqnarray} 
We have shown the numerical value of $\tilde{\lambda}$ and the above fitting functions for $\delta=1, 2$ in the Figure. \ref{figy}.  
\noindent Substituting the above expression in eq. (\ref{new21}) and the definition of the dimensionless parameter $h$, we finally obtain the critical temperature in terms of charge density in the following form
\begin{eqnarray}
	T_c &\approx& \frac{3}{4\pi}\frac{\sqrt{q\rho}}{\sqrt{ 1.09 + 0.784 \frac{\alpha_{l1}}{a\sqrt{q \rho}} + 0.987 \frac{\alpha^2_{l1}}{a^2q \rho}}} ~~~~~\text{for}~~~ \delta=1 \\
	T_c &\approx& \frac{3}{4\pi}\frac{\sqrt{q\rho}}{\sqrt{ 4.069 + 1.591 \frac{\alpha_{l1}}{a\sqrt{q \rho}} + 0.979 \frac{\alpha^2_{l1}}{a^2q \rho}}} ~~~~~\text{for}~~~ \delta=2 ~.
\end{eqnarray}
\noindent For $s$-wave holographic superconductors, the value of $\alpha_{01}=0$ since $\alpha=0$. From the above expression, 
we recover  the critical temperature for $s$-wave holographic model, which are $T_c = 0.225 \sqrt{q\rho}$ and $T_c = 0.117 \sqrt{q\rho}$ for $\delta=\delta_{-}=1$ and $\delta=\delta_{+}=2$  respectively. We present the critical temperature as function of charge density for $ l=0 (s\text{-wave}), l=1 (p\text{-wave}), l=2 (d\text{-wave})$ states of the spin-$0$ field model in  Figure \ref{fig100} for $q=1$. The $p$-wave and $d$-wave states are excited states in this model. 

\noindent Using eq.(\ref{eq36}) and eq.(\ref{eq:11}), we can now write the matter field solution as
\begin{eqnarray}
	\psi(z, u, \theta)= \sqrt{\epsilon}\psi^{(1)}(z, u, \theta)&=&\sqrt{\epsilon} \Psi^{(1)}(z) \mathcal{R}(u, \theta) \nonumber \\
	\Rightarrow \psi(z, u, \theta)&=& \mathcal{C}_s \sqrt{\epsilon} \mathcal{H}(u, \theta) \mathcal{K}(z)
	\label{afy1}
\end{eqnarray}
where $ \mathcal{H}(u,\theta)=J_l\left(\frac{\alpha_{l1}}{a}u\right)e^{il\theta} $ and $ \mathcal{K}(z)=\frac{z^{\delta}}{r_h^{\delta}}(1-\tilde{\alpha}z^2)$. Using the above expression and the gauge/gravity duality, the condensation operator of the boundary theory yields
\begin{eqnarray}
	\langle\mathcal{O}\rangle = \mathcal{C}_s  \mathcal{H}(u, \theta)\sqrt{1-\frac{T}{T_c}}=\mathcal{C}_s J_l\left(\frac{\alpha_{l1}}{a}u\right)e^{il\theta}\sqrt{1-\frac{T}{T_c}} ~.
	\label{afy3}
\end{eqnarray}
From this expression, we can identify the temperature dependent part of the condensation operator as $\langle\tilde{\mathcal{O}}\rangle =\mathcal{C}_s\sqrt{1-\frac{T}{T_c}}$. To determine the integration constant $\mathcal{C}_s$ in the above equation, we need to solve the first order gauge field equation (\ref{gfenw1}) which reads 
\begin{eqnarray}
	\partial^2_{z}A_t^{(1)}(z, u, \theta) + \frac{\nabla^2_{(2)}A_t^{(1)} (z, u, \theta)}{r^2_h g(z)} = 2\mathcal{C}_s^2 q^2 J_l^2\left(\frac{\alpha_{l1}}{a}u\right) \frac{\mathcal{K}^2(z)}{z^2g(z)}   \mathcal{A}^{(0)}_t(z) ~.
\end{eqnarray} 
Since the first order gauge field equation is not separable, it is difficult to solve the above equation. Therefore, it is not possible to determine the integration constant $\mathcal{C}_s$ here. Although the amplitude of the gap function in angle dependent scalar field model is not possible to determine, we can write the momentum dependent gap structure part without amplitude which is coming from the different wave states in momentum space in the scalar field model. Using two dimensional Fourier transformation of eq.(\ref{afy3}), we obtain the gap function for different wave states (using eq.(\ref{eq41}))
\begin{eqnarray}
	\Delta_{k} = \Delta(T) \mathcal{I}(k,\theta)= \mathcal{C}_s\sqrt{1-\frac{T}{T_c}}  \times \left\{ 
	\begin{array}{ c l }
		2\pi a^2 ~\frac{J_1(ka)}{ka} & \quad \textrm{for } l = 0 \\
		2\pi a^2  (-i)^l e^{il\theta} ~\alpha_{l1}J_{l-1}(\alpha_{l1})\frac{  J_l(k a)}{k^2 a^2- \alpha_{l1}^2}                & \quad \textrm{otherwise} 
	\end{array}
	\right. ~~~~~
	\label{eq84}
\end{eqnarray}
where the amplitude $C_s$ is undetermined here. We can compare the gap structure $\mathcal{I}(k,\theta)$ for excited $p$-wave state in the scalar field model with ground state in vector field model.

\end{document}